# Monitoring fetal electroencephalogram intrapartum: a systematic literature review

**Aude Castel[1], Yael S. Frank[2], John Feltner[3], Floyd Karp[4], Catherine M. Albright[2], Martin G. Frasch[2,5]***

[1]*Dept. of Clinical Sciences, Faculty of Veterinary Medicine, Université de Montréal, QC, Canada*
[2]*Dept. of Obstetrics and Gynaecology, University of Washington, Seattle, WA, USA*
[3]*Dept. of Pediatrics, University of Washington, Seattle, WA, USA*
[4]*School of Pharmacy, University of Washington, Seattle, WA, USA*
[5]*Center on Human Development and Disability, University of Washington, Seattle, WA, USA*

*** Correspondence:**

Martin G. Frasch
Department of Obstetrics and Gynecology
University of Washington
1959 NE Pacific St
Box 356460
Seattle, WA 98195
Phone: +1-206-543-5892
Fax: +1-206-543-3915

**Email:** mfrasch@uw.edu

**Abstract**

**Background**: Studies about the feasibility of monitoring fetal electroencephalogram (fEEG) during labor began in the early 1940s. By the 1970s, clear diagnostic and prognostic benefits from intrapartum fEEG monitoring were reported, but until today, this monitoring technology has remained a curiosity.

**Objectives**: Our goal was to review the studies reporting the use of fEEG including the insights from interpreting fEEG patterns in response to uterine contractions during labor. We also used the most relevant information gathered from clinical studies to provide recommendations for enrollment in the unique environment of a labor and delivery unit.

**Data sources**: PubMed.

**Eligibility criteria:** The search strategy was: ("fetus"[MeSH Terms] OR "fetus"[All Fields] OR "fetal"[All Fields]) AND ("electroencephalography"[MeSH Terms] OR "electroencephalography"[All Fields] OR "eeg"[All Fields]) AND (Clinical Trial[ptyp] AND "humans"[MeSH Terms]). Because the landscape of fEEG research has been international, we included studies in English, French, German, and Russian.

**Results**: From 256 screened studies, 40 studies were ultimately included in the qualitative analysis. We summarize and report features of fEEG which clearly show its potential to act as a direct biomarker of fetal brain health during delivery, ancillary to fetal heart rate monitoring. However, clinical prospective studies are needed to further establish the utility of fEEG monitoring intrapartum. We identified clinical study designs likely to succeed in bringing this intrapartum monitoring modality to the bedside.

**Limitations**: Despite 80 years of studies in clinical cohorts and animal models, the field of research on intrapartum fEEG is still nascent and shows great promise to augment the currently practiced electronic fetal monitoring.

**Prospero number**: CRD42020147474

## Introduction

Perinatally-acquired fetal brain injury is a major cause of long-term neurodevelopmental sequelae, and the single greatest contributor to disability worldwide,[1,2] accounting for 1/10$^{th}$ of all disability-adjusted life years.[3] Moreover, intrapartum-related death is the 2$^{nd}$ leading cause of neonatal mortality and the 3$^{rd}$ leading cause of death in children under five.[4] Thus, there is an urgent need to identify early signs of fetal distress during labor to allow timely and targeted interventions.

Fetal acidemia contributes to perinatal brain injury,[5] and is one of the most common and potentially devastating labor complications. Acidemia occurs in about 25 per 1000 live births overall and in 73 per 1000 live preterm births;[6,7] and the risk of subsequent brain injury rises 9-fold in the setting of preterm birth. These risks are even higher with additional complications, such as intraamniotic infection or intrauterine growth restriction (IUGR). Over 90% of children with perinatal brain injury, including that causing cerebral palsy, have a normal life expectancy, but many cannot fully participate in society or fulfill their developmental potential.[8]

Today, continuous fetal heart rate (FHR) monitoring is used as an indirect surrogate indicator to suspected fetal acidemia during labor and it fails at that.[9] Fetal acidemia per se is a poor proxy to fetal brain injury.[10] It is then not surprising that FHR monitoring intrapartum does not reliably predict fetal brain injury. Moreover, the fear of missing fetal distress increases the rate of cesarean delivery, with significant maternal risk.[11] About 50% of cesarean sections are deemed unnecessary.[12] Conversely, labor is sometimes allowed to proceed when current FHR monitoring technology suggests that the fetus is tolerating it, only to discover later that fetal brain damage occurred, causing a range of signs from subtle neurologic deficits to more overt conditions like cerebral palsy.

Fetal electroencephalogram monitoring intrapartum (fEEG), as a direct monitor of fetal brain activity, was a focus of clinical research as early as the 1940s[13] and into the 1970s[14] and 1990s.[15,16] Notably, Eswaran *et al.* used a regular FHR scalp electrode and a routine GE HC Corometrics FHR monitoring device to record auditory evoked brainstem potentials, i.e., *evoked* fEEG activity.[15] Due to technical limitations and the difficulty of data interpretation, this research into fEEG was not able to be adopted into clinical practice.

The goal of this article is to provide a systematic review of the current literature on intrapartum fEEG. Using the most relevant information gathered from studies on this subject, the second goal of this review is to provide recommendations in order to help ensure successful fEEG study enrollment in the unique environment of a labor and delivery (L&D) unit.

## Methods

The methods for searching and analyzing the relevant literature and for data extraction followed recommendations from the Preferred Reporting Items for Systematic Reviews and Meta-Analysis (PRISMA) statement. The review has been registered with the PROSPERO database under the number CRD42020147474.

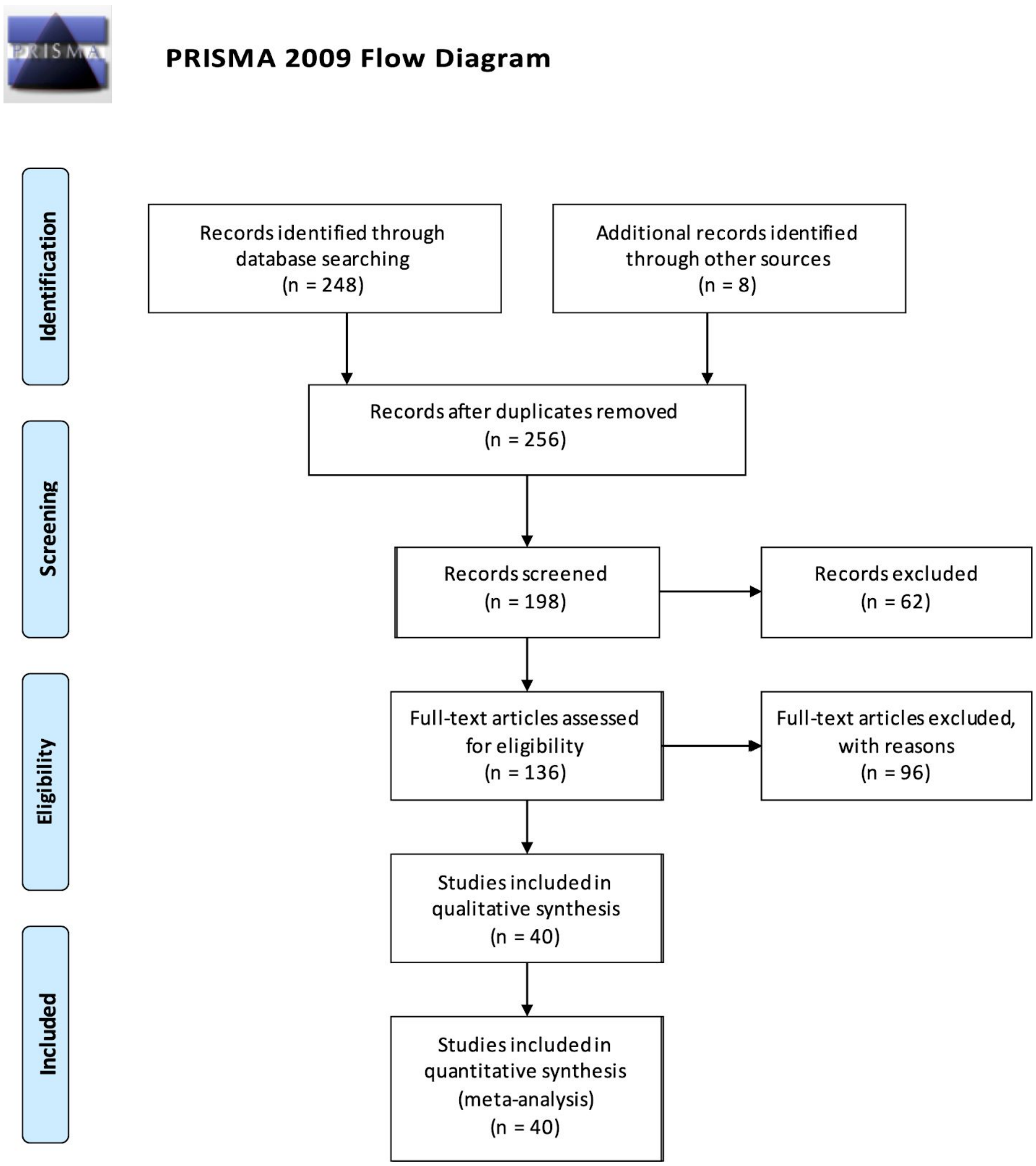

**Figure 1:** PRISMA flow diagram summarizing the study selection process and the number of studies ultimately deemed eligible to be included in the meta-analysis.

We conducted a literature search in the database PubMed covering all dates and using the following keywords: ("fetus"[MeSH Terms] OR "fetal"[MeSH Terms]) AND ("electroencephalography"[MeSH Terms] OR "electroencephalography"[All Fields] OR "eeg"[All Fields]) AND ("humans"[MeSH Terms]). All the studies retrieved with this search and available in English, French, German and Russian were screened for pertinence by the co-authors who are proficient in these languages. The literature review was completed on April 4, 2020. The eligibility criteria used to determine whether a study was included in this review or not were that the abstract and the full text described the use of EEG on the fetus during labor and provided details about how it was performed. Study selection relied on two reviewers applying the eligibility criteria and selecting studies for inclusion. More specifically, one reviewer screened all the studies and determined if they were relevant or not and the other reviewer examined all the decisions. In case of a disagreement, a joint decision was made upon discussion with the second reviewer. Non-systematic literature reviews were excluded.

The following information was extracted from each study retrieved with the above-mentioned search, and logged in a preformatted spreadsheet: the article name, authors, PubMed identification number, publication year, whether it passed screening or not (1=passed, 0=excluded), eligibility (1 = passed or 0 = excluded), the study type (human or animal model), the study size (number of subjects), the gestational age of the subjects when available, the follow-up period if applicable, the electrode configuration, sampling frequency, and monitor type. For excluded studies, the reason for its exclusion was also noted: for those excluded at screening, the reason was categorized and recorded [1=EEG not mentioned in abstract or article, 2=irrelevant]. For studies considered non-eligible, the reason was also categorized and recorded (1=No EEG monitoring, 2=No information about EEG acquisition or analysis, 3=EEG done on older children or adults and 4 = fEEG not recorded during labor or fetal magnetoencephalogram (fMEG)).

To present individual study data, quantitative data (such as gestational age) were presented as averages and standard deviations. A Prisma flow diagram was created and all the eligible studies reported in this diagram were reviewed and synthesized.

Each study was classified according to its level of evidence according to the Oxford Center for Evidence-Based medicine level of evidence.[17] Level 1 represented a systematic review of inception cohort studies, a systematic review of randomized trials, or n-of-1 trials. Level 2 represented either inception cohort studies, individual cross-sectional studies with the consistently applied reference standard and blinding, randomized controlled trials or observational study with dramatic effect. Level 3 represented non-consecutive studies, or studies without consistently applied reference standards and non-randomized controlled cohort/follow-up study. Level 4 represented a case series, case-control studies, or poor-quality prognostic cohort study. Finally, level 5 represented expert opinions without an explicit critical appraisal, expert recommendations, or first principles as well as case reports (or case series of less than or equal to 5 cases).

Finally, we summarized systematically individual study findings and used information gathered from some of the clinical studies to provide recommendations for successful enrollment in future studies in L&D units.

**Results**

A Prisma flow diagram showing the results of our database search and presenting the final number of studies included in the systematic review is shown in Figure 1.

Our initial search yielded 248 results. Eight additional studies were added following cross-referenced review bringing the total number of identified studies to 256. Of these, 34 articles were discarded because they were in a foreign language other than French, German or Russian, 23 additional articles because they were non-systematic reviews and 1 more was excluded because we could not get access to the full text. The initial screening with abstract reviewing was therefore performed on 198 studies and the number of relevant studies was further reduced to 136: 14 additional articles were excluded because there was no EEG performed; the remainder of the studies (n = 48) was excluded because they were deemed irrelevant to the subject of our review.

Of these 136 studies, 11 were excluded because they were literature reviews that were not caught in the initial screen and one was an abstract only. The full text was examined for eligibility in the remaining 124 studies.

Of these, 4 studies were excluded because fEEG was not performed as part of the experiment (only as a side test), 16 studies were excluded due to lack of information about the fEEG acquisition or analysis, 22 studies were excluded because EEG (or fMEG) was performed on the fetus but not intrapartum (5 of them were in preterm fetal sheep, 2 in preterm guinea pigs, and the rest in a preterm fetus in utero) and 40 studies were excluded because the EEG was performed on neonates after birth, or on older children (i.e., not on fetus or neonates) or adults (the mother). In an additional study, fEEG was studied just before and after labor (but not during), so it was also excluded. One last study was excluded because it was found to be a duplicate from another study written in a different language. Therefore 40 studies were ultimately included in our analysis.

A summary of the 40 eligible studies is provided in Table 1.

*Critical evaluation of the level of evidence*

Among the 40 eligible studies, none had a level of evidence of 1, 10 studies had a level 2, 11 studies had a level 3, 12 studies had a level 4 and 7 had a level 5. With our search criteria, we identified only a small number of studies with a high level of evidence (i.e., 2 or above), especially studies in humans. In particular, the older studies were mostly either the author's personal experience, case reports, or poor-quality cohort studies as fEEG was in the early experimental stages. However, these studies have the benefits of describing how the technique was developed and perfected over the years to allow determination of normal intrapartum fEEG pattern as well as recognition of patterns that could be indicative of fetal distress. We summarized below the most relevant information gathered from the 40 eligible studies on intrapartum fEEG. We divided them between studies in human fetuses and studies using animal models.

*Results of the individual studies*

Details about the condition under which the fEEG was performed and the monitoring characteristics for all eligible studies are provided in Table 2.

Studies in human subjects

The first report of fetal EEG was a case report by Lindsley (1942) who studied his own child during the 3rd trimester of his wife's pregnancy.[13] For this recording, abdominal probes were used and the tracing had a significant amount of artifacts preventing proper assessment.

Most of the eligible studies in humans date back from the 1960s-70s and more precisely originate from Rosen, Chik, and their team who are among the pioneers of fEEG recording during labor. Several of the findings described in these studies seem to overlap and are summarized below.

***Electrodes***

The use of fEEG in humans using scalp electrodes during labor was initially reported by Bernstine et al. (1955).[18] Later, Rosen and his team perfected the technique.[19] A good electrode was defined as: 1) safe to use and easily applied during labor, 2) screening out electrical artifacts such as the movement of the fetal head, maternal movements and the electrical "noise" of uterine contractions, 3) eliminating the electrical pattern of the FHR from the tracing and 4) providing EEG of a technical quality equal to that in the extrauterine environment.[20] The group tried different techniques. They initially reported the use of metal skin clips soldered to a shielded cable, coated with non-conductive plastic glue, and filed at their tip to prevent deep scalp penetration.[21] This type of electrodes was replaced by cup electrodes, initially with a platinum needle embedded in a lucid disc[20,21] (with possible skin penetration of 1-2 mm), later replaced by a central silver or platinum pin avoiding penetration of the fetal skin.[22] Although this technique seemed to provide reliable and interpretable fEEG signals, artifacts from fetal electrocardiogram (ECG) or movements of the leads remained a common occurrence and these electrodes required continuous suction to stay in place. Mann et al. (1972) described the use of a vacuum electrode similar to Rosen et al. but with a silver disc electrode used instead of their platinum needle, thus preventing puncture of the fetal scalp. The main feature of their electrode was the 100% conductivity with a silver cup, wire and plug, low resistance, good suction, and no clogging of the orifices to the vacuum source with the use of a mesh filter.[23]

In 1974, Heinrich et al. reported the use of a new intrapartum multimodal fetal monitoring device, the RFT Fetal Monitor BMT-504, that was capable of recording fEEG and tissue oxygen pressure among other parameters (ECG, pressure signals like intraamniotic pressure, temperature, heart rate) combined with either stainless steel clip electrodes or the current standard of care screw electrodes by Corometrics (USA).[24] Weller et al. (1981)[25] later described the use of a flexible electrode incorporating a guard ring surrounding the recording sites and forming the indifferent and common electrodes, with the guard ring acting as a short circuit for fECG to prevent its artifact on the EEG tracing. Suction was not needed to maintain in place this type of electrode and its pliability allowed it to be inserted through a 3 cm dilated cervix even if the two electrodes were 23 mm apart. Infrared telemetry was used to display and record fEEG, preventing power line interference, avoiding trailing leads between patient and monitoring equipment, and ensuring electrical safety. Artifacts due to the movement of leads were also prevented by incorporating the first stage of amplification in the composite assembly thus avoiding long wires carrying low-level signals. With this device, artifact-free fEEG recordings were obtained 80% of the time and uterine contraction did not affect the signal. However, in the case of a breech presentation, no fEEG could be recorded.

***Problems and limitations of the technique***

The two major problems associated with intrapartum fEEG precluding its routine use were technical issues and data interpretation. Placement of electrodes over the occipital area is the area where electrodes are most easily applied but because the occiput is a relatively quiet electrical area of the brain, the parietal area is preferred.[21] Because of the limited space, there is only a limited number of electrodes that can be placed precluding comparison of homologous areas of the brain.[21] Additionally, the moist scalp and uterine environment can attenuate potentials. Therefore, isolating the scalp from the environment by using suction allowed the recording of higher amplitude potentials.

Failure to obtain adequate EEG tracing was reported to most often occur when the signal was obscured by fECG.[20] Simultaneous recording of fEEG and fECG was shown to aid in the recognition of ECG artifacts (Figure 2).[26] Thankfully, newer electrodes were later developed to help limit the number of artifacts from fECG and movements.[25] Finally, the use of infrared telemetry to transfer the fEEG to display and recording equipment helped to prevent power line interference as previously mentioned.[25]

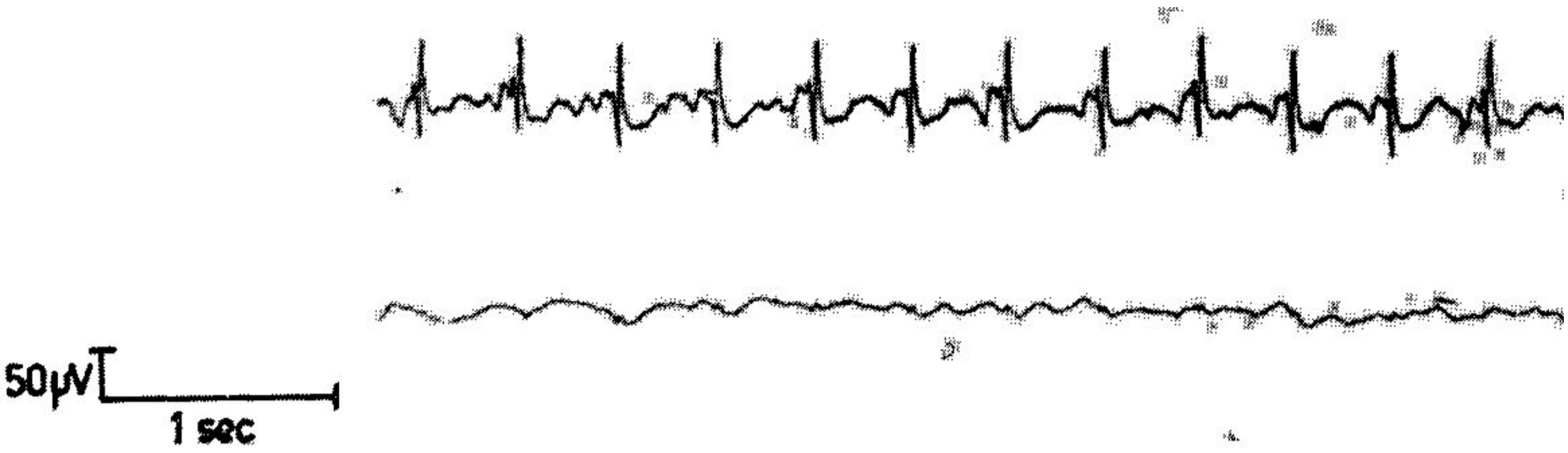

**Figure 2:** Simultaneous recording of fECG and fEEG. Artifacts from fECG effect on fEEG can be identified by recording both traces simultaneously. From[26].

Another initial limitation of the technique was the amount of information that needed to be visually interpreted. Indeed, visual interpretation had significant methodologic and interpretation bias and required certain expertise preventing routine use of fEEG as part of intrapartum monitoring. Evaluation of the value of digitized minute-to-minute and even second-to-second fluctuation in fEEG amplitude and frequency was reported by Peltzmann et al.[27]. These authors used a computer system to extrapolate the mean baseline fEEG line crosses (per 5-seconds epochs) as well as the mean integrated fEEG amplitude and presented the data in graphs with plotted point corresponding to the calculated mean by 5-seconds epochs. This represented the first steps toward simplification and standardization of fEEG signal analysis.

At the same time, computer algorithms were created to facilitate fEEG signal interpretation and standardize their evaluation.[28] A computer program was developed to help fEEG analysis by replacing the cumbersome visual analysis in an effort to integrate fEEG in computer-assisted intrapartum data management and was shown to provide 85-95% consistency with visual interpretation.[28] This program classified fEEG patterns as Low Voltage Irregular, Mixed, High Voltage Slow, Trace Alternant, Voltage Depression, Isoelectricity, and Artifact.

Visual processing of the fEEG in the form of the spectral display as an adjunct to digital analytic technique to reduce the ambiguity in fEEG interpretation was initially described by Peltzmann et al.[29] Years later, Kurz et al (1981) described the use of spectral power analysis performed in 30 s intervals with the results plotted continually over the course of the entire observation in waterfall style.[30] The authors suggested that continuous fEEG spectral power plotting helped detect artifacts on the fly which still occur and must be dealt with during the interpretation of the fEEG patterns.

Thaler et al. (2000)[16] reported the use of real-time spectral analysis to monitor fEEG during labor as more objective analysis of fEEG signal. Real-time Fast Fourier Transform algorithm allowed the representation of the EEG signals in terms of the relative power of the various frequencies of which it is composed. These frequencies were then displayed by using a density spectral array technique which helps visualize the contribution of each frequency band to the overall power spectrum: delta (0.3 to 3 Hz), theta (4 to 7 Hz), alpha (8 to 11 Hz), sigma (12 to 14 Hz) and beta (15 to 32 Hz). The brightness of a given pixel represented the relative power present at the corresponding frequency element in the fEEG. A spectral time record appeared as a black and white or grayscale image in which a given spectrum would take up only a single row of pixels. In addition, the display of the Spectral Edge Frequency (SEF) indicated the highest dominant frequency of the fEEG signal (i.e., the frequency below which 90% of the spectral power resides).

More recently, our team created automated algorithms for unsupervised fEEG-FHR monitoring and for the detection of fEEG-FHR patterns pathognomonic of adaptive brain shut-down as an early response to incipient acidemia and cardiovascular decompensation.[31]

### *Intrapartum EEG findings*

The early studies described the fEEG signal observed during labor under different conditions and while most of them were initially just observations, they allowed to gain the experience needed to determine what a normal fEEG during labor should look like and what should be interpreted as abnormal.[14,20,21,23,32–36]

To document the fEEG activity, the technique of evoked response can be used[20], although results can be quite unpredictable with significant artifacts.[37]

A summary of fEEG findings associated with normal labor, abnormal labor, and following drug administration is presented in Table 3.

1. *EEG findings during normal labor*

During labor, a low voltage baseline pattern is noted.[21] The study of 14 acceptable fEEG revealed that the voltages varied from 5 to 50 µV/cm and the wave frequencies were found between 1 and 25 Hz.[20] A small change in electrical activity was noted after delivery and rarely low voltage (20 µV), faster (8 per second) waves compared to the fEEG trace seen 30 s after delivery, and not seen before were observed after the umbilical cord was clamped. On most tracings, the electrical activity before and after the first breath and before and after the cord was clamped did not appear to change abruptly.[21,38] As the recording continued, the electrical activity slowly increased in voltage and approached that seen in similar brain regions in neonates. About 5 min after delivery, the tracing could not be distinguished from the tracing of alert neonates several hours old. Rosen et al (1965) concluded that fEEG activity recorded early in labor has a baseline pattern similar to that of the alert neonate.[21] Studies of 125 additional fEEG by the same team confirmed that the fEEG patterns observed during normal labor were similar to those present in neonates of the same weight. The wave frequencies varied between 0.5 and 25 Hz with the predominant frequencies in the 2.5 - 5 Hz.[19] Similarly, Chachava et al (1969)[39] reported fEEG findings during 20 normal labors and found that healthy (physiological) fEEG was characterized by low-amplitude waves of 0.04 - 2 seconds duration which the authors note was within the range of the reported spectrum of antenatal fEEG frequencies observed (0.5-30 Hz according to Humar & Jawinen as well as according to Bernstine & Borkowski).[18,40] They reported an amplitude of 10 - 30 µV with the observation of alpha, beta, theta, and delta waves.

Hopp et al (1972) reported simultaneous acquisition of fEEG and cardiotocogram (CTG) during labor using 3 scalp electrodes (2 biparietal and 1 midline). Normal fEEG was characterized by an amplitude ranging between 10 and 70 µV with high variability in frequencies ranging between 2 – 20 Hz. They also concluded that fEEG and neonatal EEG are basically identical and could not be differentiated from each other (Figure 3). Borgstedt et al. (1975) also reported normal fEEG showing wave frequencies of 0.5 to 25 Hz with an amplitude generally between 50 and 100 µV/cm, similar to neonatal EEG.[22] Three studies from the same group reported similar findings with intrapartum fEEG showing alternance of active and quiet sleep phase similar to neonates.[33,34,38]

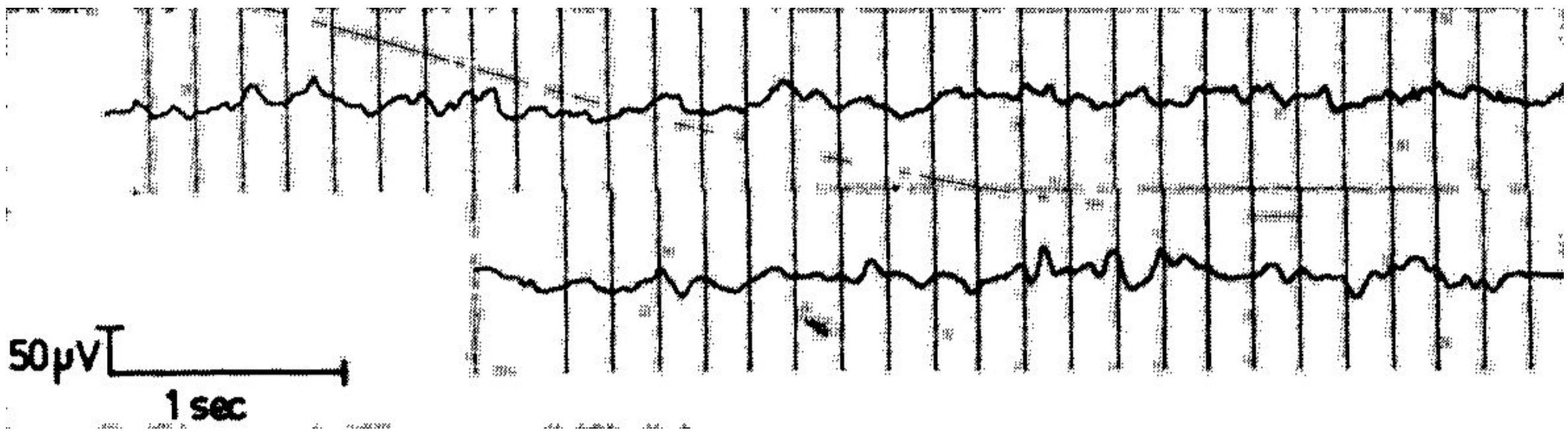

**Figure 3:** Intra and post-partum fetal/neonatal EEG recordings showing the great similarity between both traces. From Hopp (1972).[26] EEG: electroencephalogram.

In a study by Mann et al. (1972), adequate fEEG were obtained and studied in 50 patients.[23] The EEG prior to, during and following a very intense contraction (approximately 95 mmHg after oxytocin infusion) was characterized by a rhythm consisting of 1 to 3 Hz waves with an amplitude of about 40 to 75 μV with superimposed faster frequencies of 4 to 8 Hz and 10 to 30 μV. There were no significant changes in the fEEG signal during the uterine contraction and this fEEG was very similar to that of the same patient examined 18 hours after birth. The lack of influence of uterine contractions or expulsion on the fEEG signal was also documented by Chachava et al. (1972) and Challamel et al. (1974).[33,41] Similarly, fEEG recorded during the second stage of labor did not show any alteration in frequency, amplitude, and pattern despite the increase in uterine pressure associated with maternal pushing (a contraction of abdominal wall muscles).[14] Conversely, in a study using simultaneous CTG and fEEG recording under conditions of intermittent hypoxia due to uterine contractions, fEEG showed a reduction of frequency and increase of wave amplitude during contractions.[42]

During spontaneous birth, a low voltage irregular activity was noted as well as artifactual distortion of the fEEG baseline characterized by large rolling waves of almost 2 s in duration due to electrodes movements when the vertex moves rapidly and the fEEG is recorded in the microvolt range.[14] This appears to be a common problem during the birth process.

The effect of head compression associated with cephalopelvic disproportion on fetal brain activity was studied and no significant differences in fEEG findings between the group with cephalopelvic disproportion and the group without it were noted.[43]

Using real-time spectral analysis, a more objective method of fEEG assessment, Thaler et al (2000) identified two fundamental fEEG patterns in the recording: high voltage slow activity (HVSA) (quiet behavioral state) and low voltage fast activity (LVFA) (active behavioral state).[16] FHR accelerations were typically associated with periods of LVFA but there was no relationship between uterine contractions and SEF or density spectral array (DSA) (power spectrum). The 90% SEF was found to be an excellent index of cyclic EEG activity. When combining the results of the 14 fetuses, it was found that on average, LVSA was present 60.1% of the time and HVSA was present 39.9% of the time.

2. *Abnormal EEG findings*

Chachava et al (1969) first reported fEEG during complicated labor and presented the case of a baby born asphyxiated and demised within 15 min postpartum.[39] The fEEG showed fast activity around 6 Hz that was suggested to represent brain hypoxia but the changes were not considered unique and pathognomonic. High amplitude low-frequency waves were, in their experience, signs of intrapartum brain injury.

During labor, transient or persistent fEEG changes can be observed. Usually, persistent changes are considered to be abnormal if they occur between two events such as uterine contraction or expulsion efforts leading to a progressive deterioration of the fEEG activity.[44]

Evaluation of the fEEG signal associated with FHR changes revealed different situations which we summarized below.

Simultaneous recording of fEEG and CTG/FHR revealed that slow waves and frequency decrease could be observed during and shortly after uterine contractions and were seen as an expression of short-term brain ischemia due to an increase in intracranial pressure (Figure 4). The vagal stimulation inducing the early decelerations in CTG was thought to be due to an increase of intracranial pressure, but indirectly, with the primary vagus stimulation trigger being due to transient cerebral hypoperfusion during uterine contraction. Fetal bradycardia, especially during contraction-associated late decelerations, was accompanied by a reduction in fEEG waves (lower frequency) and occurrence of fEEG spike potentials.[26]

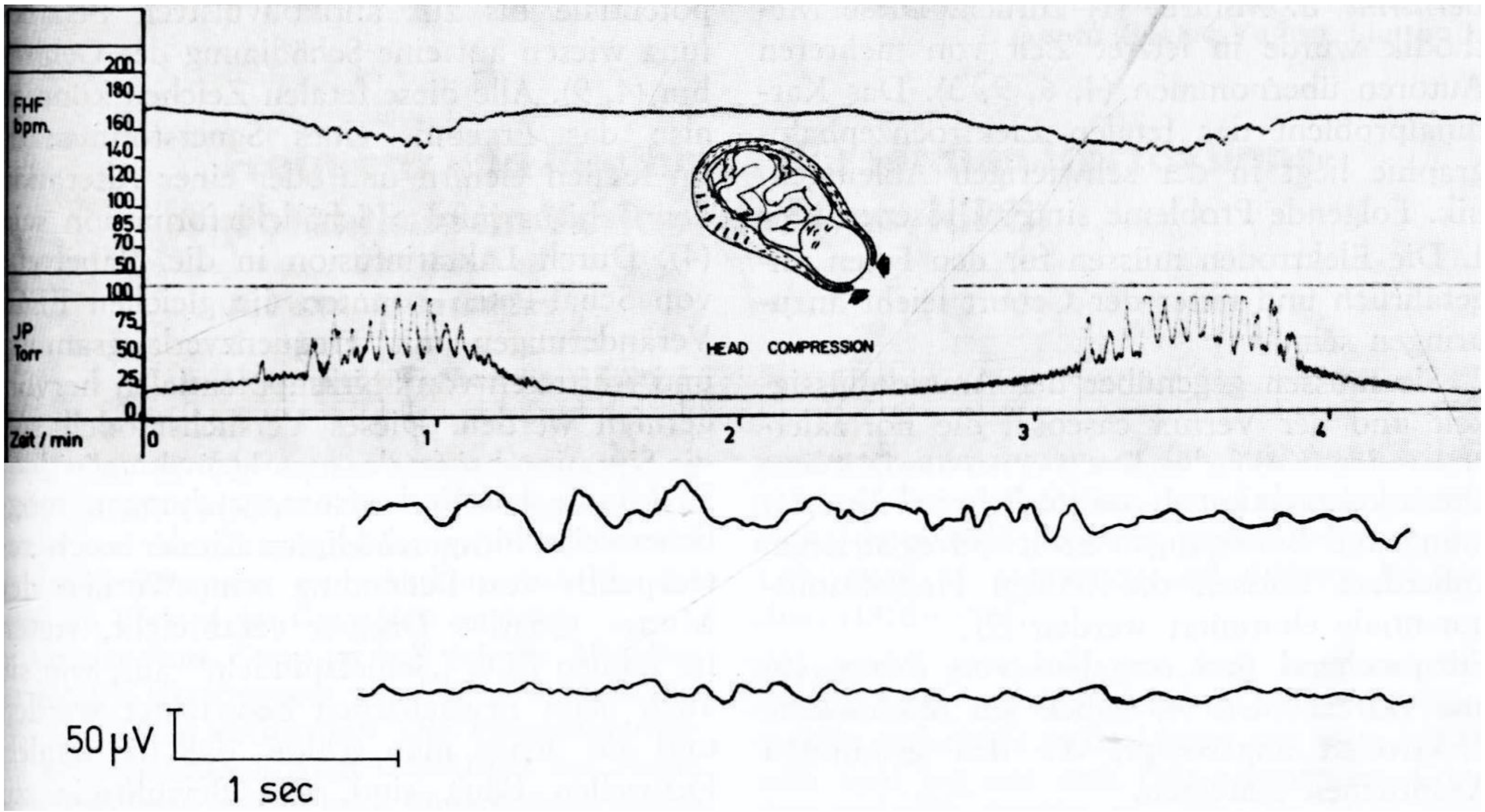

**Figure 4:** Cardiotocogram (top) and fEEG (bottom) recorded during early cardiac deceleration. The fEEG pattern represents the change during contractions with high amplitude low-frequency waves and the recovery once the contractions ceased. From .[26]
fEEG: fetal electroencephalogram.

In one study by Rosen et al[32], transient fEEG changes were noted during FHR deceleration. The fEEG appeared to lose faster rhythms, followed by a more apparent slowing. As the condition persisted, isoelectric to almost flat periods with rare bursts of fEEG were seen. Finally, a totally isoelectric interval was observed sometimes for longer than 10 s (rarely more than 30 s). As the FHR returned to its baseline rate, the reverse of this progression took place with the entire sequence from onset to return lasting from 30 s to sometimes longer than one minute. These changes were not seen with early FHR deceleration but were observed with variable decelerations and late decelerations. They were also observed during prolonged spontaneous expulsion or expulsion of a distressed infant.[33] In another study, using simultaneous CTG and fEEG recording, severe variable decelerations were also associated with waves of low amplitude and near isoelectricity and intermittent spike potentials between contractions (Figure 5).[42]

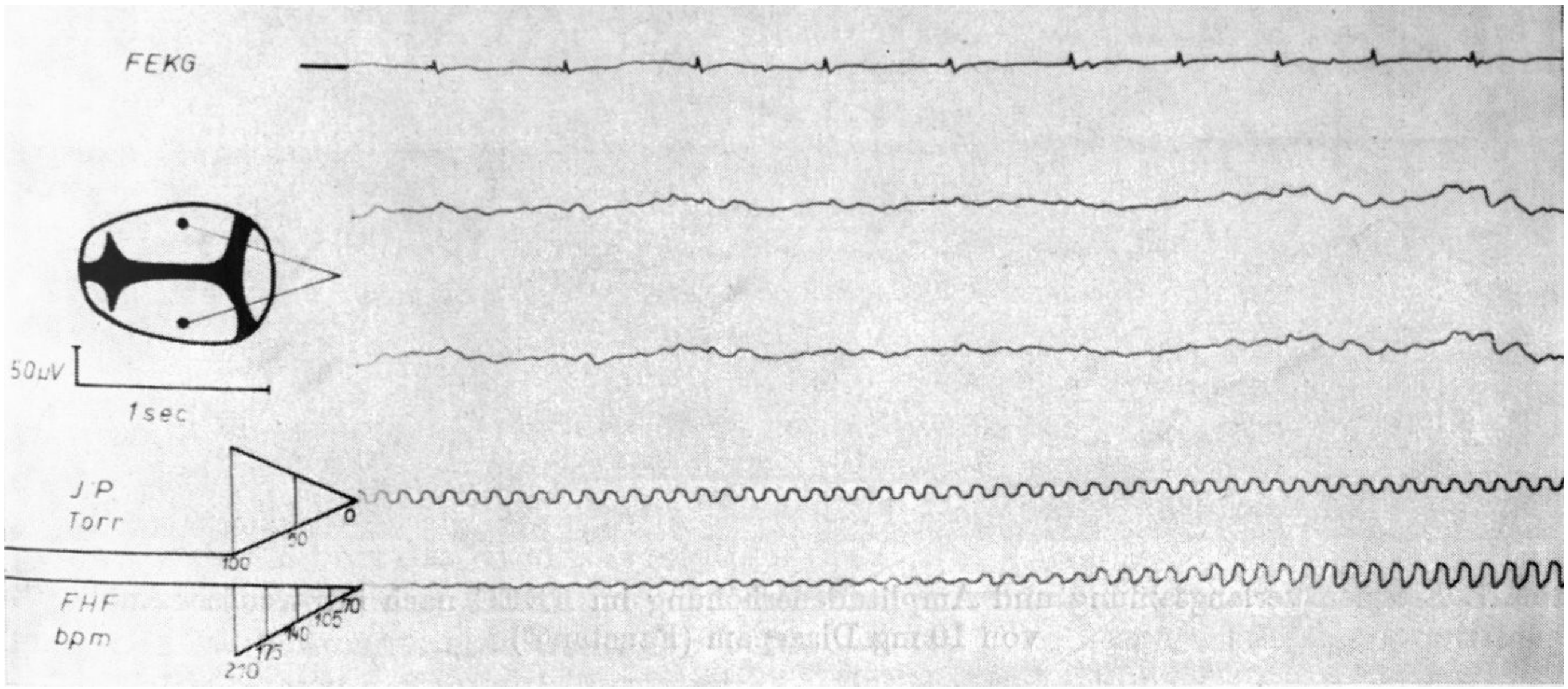

**Figure 5:** Simultaneous recording of fECG (top trace), two-channel fEEG (middle two traces), and FHR (bottom trace). This figure shows fEEG changes during severe variable deceleration. The fEEG trace shows waves of low amplitude and near isoelectricity as well as intermittent spike potentials between contractions. From [42].
fECG: fetal electrocardiogram, fEEG: fetal electroencephalogram, FHR: fetal heart rate.

Revol et al. (1977)[44] documented the fEEG changes during spontaneous expulsion and noted the following events: either no fEEG changes or transient fEEG changes not exceeding 20 s (Type I); EEG changes that disappear just before the expulsion effort (Type II); and persistent fEEG changes (Type III). During expulsion, the relationship between fEEG and FHR revealed that Type I fEEG was 85% of the time associated with transient or no FHR changes whereas Type II and Type III fEEG changes were associated with bradycardia. In their cases, early deceleration was only associated with fEEG changes with FHR below 90 bpm.[44] Spontaneous tachycardia (>160bpm) and bradycardia were both associated with fEEG changes (decreased activity and flattening of the trace).[38] However, tachycardia following atropine administration was not associated with any fEEG changes.[38] Using spectral power analysis, Kurz et al (1981) also observed a similar relationship between the degree of spectral fEEG suppression and the FHR decelerations induced by uterine contractions.[30]

Simultaneous recording of fEEG and fECG during labor after premature rupture of membranes (PROM) in a group of healthy women and a group of women with mild nephropathy, was reported by Nemeadze et al (1978).[45] Normal fEEG characteristics were 1-16 Hz, 10 - 30 µV, asynchronous, dysrhythmic activity; PROM had no significant effect on these parameters of fEEG or fECG over the course of labor. However, the immediate response of fEEG to the PROM event, identified as a period of 5 minutes, was shown to be reflective of incipient perinatal brain injury. Specifically, in some fetuses, regardless of maternal health status, PROM induced a response in fEEG characterized by pathological activity with high-amplitude slow 1-3 Hz waves, periodically acquiring a group-rhythmical character; fEEG normalized gradually within 1-3 minutes and fetuses showing this rapid recovery of fEEG all had a healthy birth. Fetuses whose EEG recovered within 3-5 minutes and belonging to mothers with mild nephropathy, however, were diagnosed with brain injury at birth. It is thus concluded that pregnancy complications, but not the PROM itself, impact the acute fEEG response to PROM and may provide valuable insights into therapeutic labor management.

The effect of forceps birth on fetal brain activity was also evaluated.[14,21,33,34,38] FEEG recorded during labor involving forceps application required placement of the two electrodes along the sagittal suture and between the fontanelles to avoid the forceps blade (compared to their placement over the parietal region for normal birth).[14] Aperiodic, 50μV, 0.5-5Hz slow waves were reported to become more apparent when forceps was applied or when the vertex was on the perineum and the mother bore down.[21] Forceps application was not associated with any changes in the fEEG signal but during traction, an almost flat tracing was observed. Tracing resembling a burst suppression pattern could also be observed in some cases.[14] Another study comparing high and low forceps extraction revealed that high forceps extraction was always associated with fEEG changes during the traction phase and was characterized by flattening of the trace returning to normal after a few seconds if the extraction was short and not too intense.[33] Repeated and prolonged tractions were associated with persistent isoelectric trace up to the birth of the child and sometimes persisting for at least 20 minutes after birth.[33,38]

A significant correlation between the development of electrocerebral silence in the fEEG during the final hour of the first stage of labor and the development of fetal acidosis at the end of the first stage of labor was reported.[43] The rapid deterioration in the fEEG occurred as the pH fell and even at preacidotic levels (pH of 7.2 to 7.25), marked changes were present with the cessation of electrical activity in the fetal brain. A significant relationship was also noted between the increasing percentage of electrocerebral silence and the development of FHR deceleration patterns during labor. In the study by Wilson (1979), different from Rosen et al (1973), early FHR deceleration was associated with prolonged silence in the fEEG. Similarly, intermittent suppression of fetal brain electrical activity during FHR decelerations induced by umbilical cord occlusions and also arising at around pH values of 7.2 was reported in fetal sheep models of human labor.[46,47]

Another fEEG study in 11 cases of fetal distress revealed a loss of fEEG variability, sometimes similar to the awake state. These changes were usually transient during events or maneuvers.[33] A decrease in fEEG amplitude and frequency has been reported with the uterine hypertonicity of hyperkinesia.[33,34] Revol and his team studied fEEG changes associated with fetal distress in 37 cases (fetal distress diagnosed with a combination of abnormal scalp pH, umbilical blood pH, and Apgar score at 1 min). In 4 additional cases, some changes in fEEG were suspicious for fetal distress. The fEEG was abnormal in 39 of these 41 cases. The 26 cases for which all the aforementioned criteria of fetal distress were present had the lowest 1 min Apgar score (between 1 and 7). In 8 of these cases, *in utero* resuscitation measures allowed improvement of biological (i.e., pH) values and FHR. However, only in 2 cases did the fEEG normalize before birth which supports a delay in fEEG recovery compared to other criteria.[44] Another study looked at the correlation between abnormal fEEG findings and the 1 minute and 5 minutes Apgar scores in high-risk cases.[48] Prolonged voltage suppression periods (below 20 μV), usually present from the beginning of the fEEG recording and persisting throughout, represented a distinctive pattern significantly correlated with a low 1 min and 5 min Apgar scores. This pattern was also correlated with the employment of postpartum resuscitative measures and with the degree of resuscitation.[48]

In a study by Hopp et al (1973), simultaneous evaluation of fEEG, fetal ECG, and CTG, during the first and second stages of labor was shown to improve the detection of the fetus at risk of brain injury.[49] They reported a series of patterns pathognomonic for abnormal fEEG: 1) extremely high voltage activity (> 80 μV), 2) extremely low voltage activity(< 10 μV), 3) spike potentials as a sign of epileptiform activity, 4) bihemispheric differences, and 5) reduction of fEEG frequency during a pathologically silent FHR pattern.[26]

During the same period, Rosen and his team also reported one major fEEG abnormality, the non-transient sharp waves defined as repetitive waves always of the same polarity, generally higher in amplitude than the surrounding fEEG and generally less than 50 ms in duration.[32] When observed, they were usually present at the onset of recording and continued throughout labor and seemed to be more frequent in children neurologically abnormal at one year of age. This observation was later confirmed by retrospective fEEG evaluations to see if the infant outcome at one year of age with regard to neurological status could be predicted.[22,50] Sharp waves that appeared in isolation and not as part of burst activity were identified as abnormal. Once again, isolated sharp waves were noted to be more frequent in newborns with abnormal neurologic findings than in those neurologically normal and were significantly associated with neurological abnormalities at one year of age.[22,50]

Furthermore, retrospective comparison of intrapartum fEEG from neurologically abnormal infants at one year of age to neurologically normal children revealed that the combination of sharp waves and low voltage did not occur in the normal population suggesting that this type of activity may indicate fetal distress requiring intervention.[32,50] To further confirm these findings, the previously described computer program developed by Chik et al.[28] was used to retrospectively evaluate artifact-free EEG of these neurologically normal and neurologically abnormal infants.[51–53] In the neurologically normal group, the mixed pattern was predominant accounting for 41.2% of the 10,511 epochs evaluated. The trace alternant pattern accounted for 32.2%, high voltage slow pattern for 21.5%, and low voltage irregular pattern for 4.4% of the patterns. Less than 0.2% showed depression or isoelectric signal. In the neurologically abnormal infants, low voltage irregular activity accounted for 17.85% of the epochs, mixed activity for 30.5%, high voltage slow activity for 18.1%, and trace alternant for 33.2%. Less than 0.2% of the epochs showed depression or isoelectric signals. The number of observed fEEG patterns in abnormal cases was significantly different from normal cases. The relative frequency of the low voltage irregular pattern was increased with a decrease in mixed and high voltage slow patterns. The mean relative frequency of low voltage irregular pattern was significantly greater in the 1 min lower Apgar score (less than 9). Therefore, low voltage irregular patterns were shown to occur more frequently in the neurologically abnormal group (compared to the neurologically normal group). The same group used a computer-interpreted EEG to try to predict the infant neurological outcome at one year. Using fEEG patterns alone (by looking at the relative frequency of low voltage irregular, high voltage slow, mixed, and trace alternant patterns), almost two-thirds of the neurologically normal infants and of the abnormal infants were correctly classified. Using intrapartum fEEG and FHR patterns simultaneously provided slightly better results to predict neurologically normal infants but gave the same results for the neurologically abnormal ones. Combining intrapartum data with postpartum data, including 1 minute, 5 minute Apgar scores, and neonatal neurologic examinations, about 80% of the infants were correctly classified (Chik 1977).[53,54] These results show that combining multiple methods of peripartum fetal monitoring allows better detection of fetal distress that could affect long term neurological outcome.

3. *Effect of drugs*

Six studies reported their observations of fEEG following maternal general anesthesia with different drugs and described some characteristic changes.[19,27,33,34,38,42]

FEEG recorded following maternal anesthesia with alfatesine at a continuous rate infusion (CRI) showed changes between 1 to 11 min (mean 3.5 min) following the beginning of the CRI. Initially, theta waves occurring in clusters altering the baseline rhythm were noted. These fEEG changes eventually disappeared to the point of reaching a discontinuous aspect with alternance of theta wave clusters and isoelectric state. Electrical silence could also be observed. Theta activity was noted to persist for about 30 min after birth. The fEEG baseline activity reappeared about 40 min after birth with the persistence of occasional theta activity during different vigilance states associated with anesthesia. FEEG changes were more pronounced if fetal distress was also present.

The effect of ketamine on fEEG showed similar changes with sharp theta activity on an initially normal baseline with a progression to fewer waves and flattening of the trace to the point of isoelectricity with occasional bursts of theta activity approximately within 3 minutes following drug administration.[38,44] Barbiturates such as sodium thiopental were associated with the more significant changes with long periods of isoelectric traces.[33,38] Meperidine and diazepam were not found to be associated with any fEEG changes in a very small case series.[27] Conversely, meperidine was associated with early fEEG changes characterized by a transient increase in delta and theta wave frequencies (2.5-5 Hz), about 50 μV in amplitude, first seen between 1 and 2 minutes after intravenous injection of the drug followed by a trace-alternant-like pattern of bursty activity within 5 min after the mother was given the medication.[19] This pattern could last as long as 2 h after the injection. These results suggested a rapid transfer of the drug from the mother to the fetus. As the time interval after injection increased, the presence of faster, lower voltage forms (5-10 μV, 15-25 Hz) in the beta range would become more obvious.[19] The effect of the administration of 50 mg of pethidine revealed a reduction of amplitude and frequency of fEEG activity about 1 minute after the injection.[42] These changes were more pronounced at 4 min post-injection. At 6 min post-injection, resynchronization was observed. These effects persisted for 25 min and fEEG normalized more or less within 105 min post-injection.

Minor fEEG changes were noted with local anesthesia and were characterized by high-frequency rhythms with clusters of rhythmic theta waves.[33] FEEG changes associated with anesthetic persisted for 1 to 3 days after birth. In two very small case studies, paracervical block with 1% mepivacaine was associated with a decrease in fEEG amplitude with a questionable effect on the frequency.[27,29] Caudal or paracervical carbocaine administration was shown to produce transient pattern changes consisting of an increase in higher voltage (50 μV/cm) bursty waves (15-25 Hz).[19] In the presence of penthrane, a trace alternant picture persisted while the gas was being administered during the terminal stages of labor.[19] Finally, the effect of diazepam injection (10 mg) on fEEG was also reported. The fEEG frequency decreased within 30 min post-injection and the amplitude increased to 80 μV with normalization of neonatal EEG recorded 40 min after the injection.[42]

Of interest, the persistence of all recording voltages below 20 μV with prolonged intervals of isoelectricity, described as low voltage tracing, was observed in less mature infants in the presence of analgesic medications.[32] This pattern was associated with an initially normal amplitude and the pattern of recording changing to persistent low voltage with prolonged periods of isoelectricity.

Finally, in a study comparing fEEG before and after oxygen ($O_2$) administration by mask to 20 mothers during labor, it was shown that $O_2$ administration caused fEEG changes within 1min30s to 2min after initiation of $O_2$ characterized by a progressive increase in amplitude and frequency of the waves (from 1-5 Hz to 8-12 Hz) reaching a maximum at 7-8 minutes followed by a decrease in the activity of the trace to return to baseline activity after 12-15 minutes in half of the cases.[55]

Studies using animal models

All the animal studies deemed eligible used a fetal sheep model. Because of the similarities between ovine and human fetal physiology,[56] this species is considered a reliable model to study fetal cerebral development.[57] First, the sheep fetus displays cerebral hemodynamics similar to that in humans. Second, the sheep fetal cardiovascular and EEG data can be derived in the unanesthetized state. Third, similar to the human fetus[58–61], the sheep fetus displays a very limited range of cerebral autoregulation under normal conditions and they both have a pressure-passive cerebral circulation when subjected to systemic hypoxia and the associated hypotension.[62,6364–66] Such hypotensive response is amplified in chronically hypoxic pregnancies, such as with IUGR, where fetal myocardial glycogen reserves are more rapidly depleted under conditions of umbilical cord occlusions (UCO).[67,68]

The 6 studies selected used transient UCO mimicking what can happen during labor with uterine contractions and therefore represent a good model compromise to study intrapartum fetal distress and fEEG. Some of these studies recorded fetal electrocorticogram (fECoG) where electrodes are placed directly on the dura for optimal signal quality by comparison with fEEG where electrodes are sewed into fetal sheep's skin.

De Haan et al. 1997 recorded fECoG, allowing to record brain electrical activity similar to fEEG in sheep fetuses following repeated UCO of different duration (1 min every 2.5 min or 2 min every 5 minutes) compared to sham controls.[69] During the occlusions, there was a progressive fall in fEcoG intensity, more pronounced in the group with the longer UCO. FEcoG activity at the final occlusion and recovery to normal sleep cycling patterns were similar in the two UCO groups. A fall in SEF during UCO followed by rapid normalization during recovery was similar in the two asphyxiated groups. Two characteristic patterns of electrophysiologic changes were noted. In the baseline period, there was normal sleep cycling characterized by an alternation of high voltage and low voltage fEcoG activity. During the occlusions, the fECoG intensity decreased to eventually reach a trough at the final occlusion and recovered thereafter. In fetuses that subsequently developed only selective neuronal loss as assessed on histologic evaluation, the fECoG rapidly recovered, associated with very little epileptiform or spike activity. Conversely, fECoG tracing indicating more epileptiform activity was seen in the fetuses with most extensive neurological damage on histology and the fECoG recovery was slower in the more severely damaged fetuses. In comparison, sham fetuses showed no changes in fECoG activity (and at postmortem evaluation). Despite a similar frequency of the asphyxia periods, the longer episodes of cord occlusion appeared to have a greater initial effect on the fEcoG with significantly more epileptiform and spike activity than the shorter one reflecting the cumulative effect of intermittent ischemia with longer hypotensive periods on fECoG (and consequently fEEG) and brain injury.

Thorngren-Jerneck et al. (2001) also reported the effect of UCO on fEEG and compared the fEEG signal of 3 groups: one subjected to total UCO until cardiac arrest, one sham control group and one healthy control group.[70] The fEEG became rapidly flat during the cord occlusion in all lambs subjected to UCO and remained isoelectric during the 4 h after delivery. Conversely, the fEEG was "normal," i.e., showing continuous activity with mixed frequencies, in sham and healthy controls during the 4 h after delivery. Using positron emission tomography, they also demonstrated that global cerebral metabolic rate was significantly reduced 4 h after fetal asphyxia induced by UCO. Their findings suggest that prolonged isoelectricity identified on EEG after birth is an indication of severe fetal distress and that a reduction in the brain's metabolic rate represents an early indicator of global hypoxic cerebral ischemia.

In another study by Kaneko et al. (2003)[71], fetal sheep were subjected to UCO without regard to the electrocortical state activity every 90 minutes, and over 6 hours (for a total of four UCOs). The fECoG was monitored continuously and assessed by visual analysis into periods of high voltage (>100 µV) and low voltage (<50 µV). Following UCO, an indeterminate electrocortical pattern became apparent with initially lower than baseline electrocortical state and then gradually increasing toward a high voltage electrocortical state but with no evident cycling. The fetal electrocortical activity was disrupted markedly by 4 minutes of UCO, with an abrupt flattening of the fECoG. With the release of the cord occluder, the fECoG amplitude increased steadily over several minutes. These results show that UCO resulted in a progressive decrease in fECoG amplitude with most animals showing a flat ECoG by 90 s but with rapid recovery in voltage amplitude after the release of the occluder. These results are similar to what has been reported in humans following severe cardiac deceleration.[32,43]

Our team conducted several studies using a fetal sheep model of human labor and showed that certain changes in fEEG and fECoG accurately predicted severe acidemia during labor with sufficient lead-time to potentially intervene and perform a cesarean section.[31,46,4772] This method was shown to have a positive predictive value (PPV) of 70% and a negative predictive value (NPV) of 100%.[31] Using ECoG and EEG recordings, we identified pathognomonic changes in fetal electrocortical activity predictive of cardiovascular decompensation and severe acidemia allowing early (~60 min) recognition of a critical situation and giving sufficient time to perform an emergency cesarean section.[4631]

The utility of joint fEEG-FHR monitoring is based on the consistent emergence of synchronized UCO-triggered blood pressure, and fEEG-FHR changes, prior to reaching a severe level of fetal acidemia where brain injury might occur. The fetal blood pressure showed a pathological hypotensive behavior concomitant with the fEEG-FHR changes during FHR decelerations (Figure 6).[31] These changes are thought to be due to adaptive brain shut-down, triggered at a pH of about 7.20. Of note, Yumoto *et al.* also reported a pH of 7.20 to be the critical value, below which fetal myocardial contractility begins to decrease.[73] Adaptive brain shutdown prevents the brain from passing from upper to lower ischemic flow thresholds.[74,75] When the fetal brain blood flow falls beneath the lower ischemic flow threshold, permanent neurological injury occurs.[46]

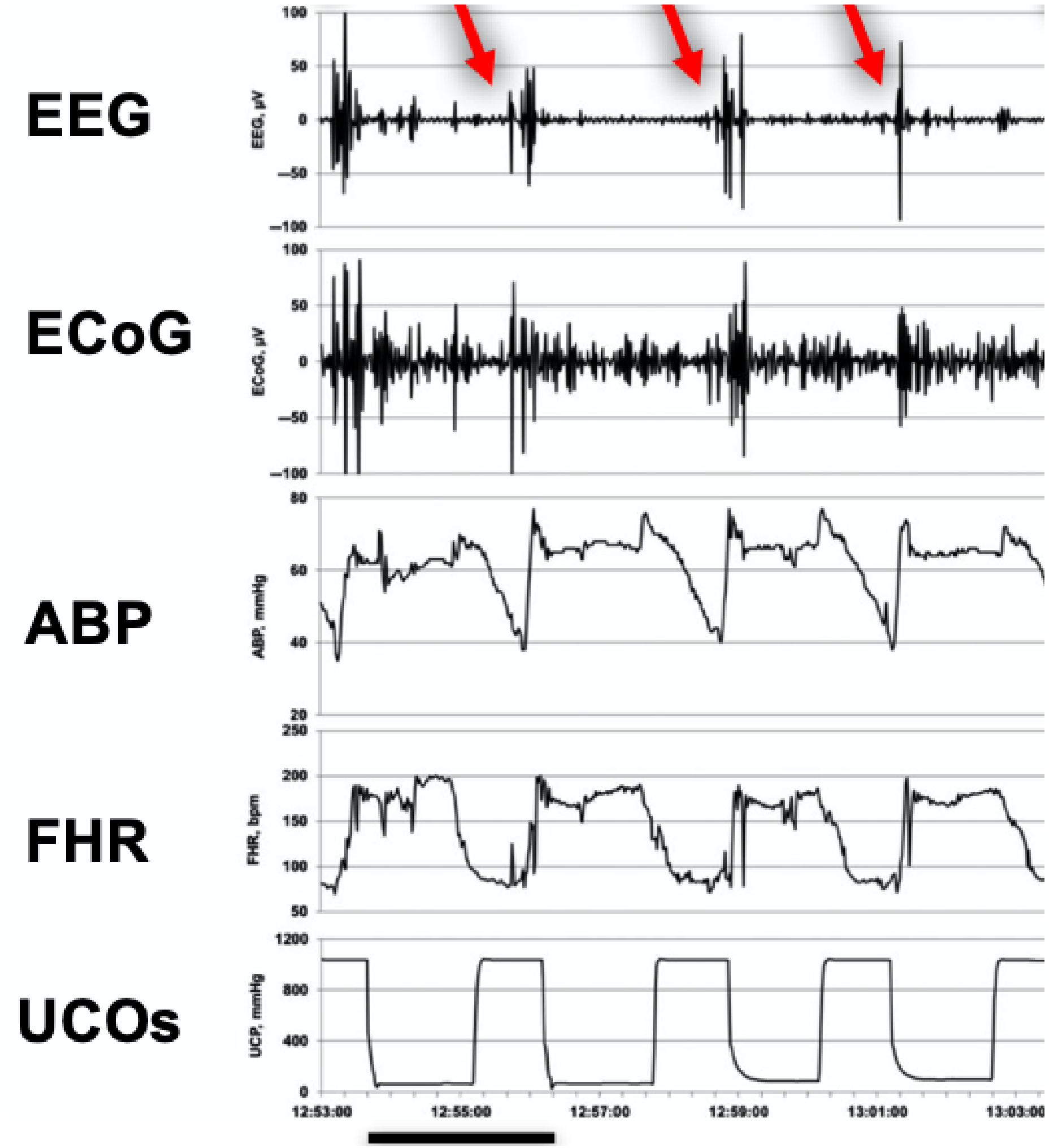

**Figure 6: Emergence of EEG-FHR pattern in a fetal sheep model.** A representative 10 min recording made during the early stage of severe umbilical cord occlusions (UCOs) at a pH of about 7.2 and about 60 min prior to pH dropping to less than 7.00 indicated cardiovascular decompensation (hypotensive fetal systemic arterial blood pressure; ABP) in response to FHR deceleration triggered by UCO. It shows the pathognomonic fEEG pattern (black bar = 2.5 min). Red arrows indicate the pathognomonic burst-like EEG activity correlated in time to the FHR decelerations and pathological ABP decreases during the UCOs. UCOs continued until pH < 7.00 was reached in each fetus (about 4 hours). Fetal arterial blood samples were taken each 20 min. This timing corresponds to pH of 7.20 seen in 20% of births.[76] From [31] EEG = electroencephalogram, µV; ECoG = electrocorticogram, µV;

ABP = fetal systemic arterial blood pressure, mmHg; FHR = fetal heart rate, bpm; UCOs = umbilical cord occlusions, mmHg (rise in occlusion pressure corresponds to an UCO).

Finally, the chronically instrumented non-anesthetized fetal sheep model with UCO was also used to study the presence of epileptiform activity during rewarming from moderate hypothermia, one of the undesirable outcomes associated with this common therapy for HIE.[77] Cerebral ischemia was induced by transient carotid occlusion corroborated by the onset of an isoelectric fEEG signal within 30 s of occlusion. Sheep fetuses were randomized to either cooling or sham cooling starting at 6 h after ischemia and continued until 72 h. Rebound electrical seizure events were observed in about 50% of the cooled animal and 7% of the sham-cooled animals. These results demonstrated that following a severe ischemic insult treated with moderate cerebral hypothermia, rapid rewarming was associated with a significant but transient increase in EEG-defined seizure events.

Taken together, these findings, similar to what has been reported in humans, further emphasize the relevance of the fetal sheep model to study labor-associated fetal and neonatal cerebral ischemia and help develop and validate new monitoring and therapeutic interventions.

*Synthesis of results*

The systematic analysis of the literature on intrapartum fEEG remains relatively scarce and somewhat outdated with a lot of redundant or confirmatory information. However, studies in human patients, corroborated by studies using animal models suggest that this monitoring modality can provide valuable information about fetal brain activity that significantly influences and predicts the neurological development of the newborn.[22,31,51,52]

One of the key features of fEEG is the ability to potentially detect cerebral activity changes secondary to fetal distress sooner than with evaluation of FHR alone and more continuously than by relying only on fetal scalp blood pH, a technique hardly used in the modern practice.[43,46,77] If the technical difficulties associated with electrode placements have been mostly removed,[25] the problem of objective data analysis and interpretation, although improved by the use of computer algorithms[28,31,51,52] and spectral analysis[16,27,29,30] remains a significant limiting factor in democratizing the use of intrapartum fEEG as part of the routine labor monitoring. Despite compelling evidence that joint fEEG and FHR monitoring and detection of pathognomonic patterns associated with fetal distress are key features of intrapartum fetal health assessment, the development of methods allowing unsupervised monitoring of these two variables without requiring a high level of expertise, remains in its infancy.

Furthermore, as most human studies were either retrospective cohort studies or case series, more clinical prospective studies are needed to further establish the utility of fEEG monitoring intrapartum. We identified clinical study designs likely to succeed in bringing this monitoring modality as a bedside test in the unique setting of L&D and will be discussing them below.

*Risk of bias across studies*

To limit the risk of bias for each individual study, we ought to assess the studies at the outcome level. However, because the majority of the eligible studies, particularly the ones in humans, reported mainly descriptive findings, this turned out to be extremely challenging. Indeed, a lot of these studies just described fEEG traces of selected cases.[20,21,25,27,29,32–35,37,78]. In fact, only 4 studies analyzed the fEEG in relation to the outcome at one year and are from the same group (with the same cohort for all but one study).[22,50–52]

We did try to limit bias in study selection by not just including studies in English, but also those in French, German and Russian which added 30 studies to the screening process with 14 ultimately deemed eligible.

## Discussion

*Summary of evidence*

The review of the aforementioned eligible studies allowed us to establish some key-points about intrapartum fEEG. A normal baseline intrapartum fEEG activity was reported in several studies with evidence of alternance of sleep/wake states including two types of sleep behaviors (active and quiet).[21,23,35,38,7920] This "normal" intrapartum fEEG activity was similar to that of a newborn of the same age and same birth weight. Similarly, several studies identified patterns suggestive of fetal distress. Drugs, in particular, if given systemically, were shown to influence fEEG activity. Finally, a correlation was established between fEEG activity, FHR deceleration, Apgar scores (1 minute and 5 minutes), and these factors were shown to be useful to predict the neurological outcome of the infants at one year of age. Animal studies using fetal sheep models and UCO were able to reproduce some of the abnormalities associated with fetal distress and showed that fEEG activity assessment could be a useful monitoring tool to help detect abnormal fetal brain activity associated with intrapartum complications.

The "normal" intrapartum fEEG activity was reported by several studies as a low voltage baseline pattern that varies from 5 to 50 µV per cm, with waves frequencies between 0.5 and 25 Hz. A predominant theta activity or an alternance of delta and theta activity were observed.[23,35] It is interesting to note that none of these early studies reported fEEG amplitude above 200 µV. We were able to record human intrapartum fEEG with a fetal scalp electrode with amplitudes around 400 µV. The data was acquired at 1000 Hz. In this case, the amplitude of the raw signal is about twice the reported maximum of about 200 µV.[23,7920]. It is possible that this high amplitude is the result of the effect of diazepam administration as reported by Hopp (1976) and Khopp (1977).[42,80] It is also possible that the older technologies and the filters used about 50 years ago might have prevented the capture of the intermittent faster waves with higher amplitude. This assumption is supported by the following: if we filter our recording similarly (i.e., 0.5 - 12 Hz), the fEEG tracing resembles more what these studies presented (amplitude below 200 µV) (Figure 7). A distinctive high-/low-frequency behavioral state pattern during the first stage of labor is seen as an alternance of 10 Hz and 2 Hz fEEG activity (Figure 7, TOP). It would, therefore, be interesting to repeat some of these older studies with the newest digital EEG technology.

Different studies have identified distinctive patterns suggestive of fetal distress and potentially associated with an abnormal outcome at one year of age. Particularly, sharp waves and long voltage depression were both reported to be more commonly identified in cases of fetal distress and neurologically abnormal children at one year of age.[14,22,48,51,52]

The effect of intrapartum drug administration to the mother (for analgesia or anesthesia) was also reported in different studies and appeared more significant if the drug was given systemically (in comparison to local anesthesia).[19,27,29,33,34,38]

One of the current limitations for routine use of fEEG monitoring remains the expertise required to read and interpret the tracing. Computer algorithms and methods to digitize the fEEG signal (including spectral analysis) have been developed but have remained experimental, failing to be translated to day-to-day practice.[16,28,31] Computer-assisted fEEG reading and interpretation should be further developed to help democratize this tool allowing its routine use in an L&D unit.

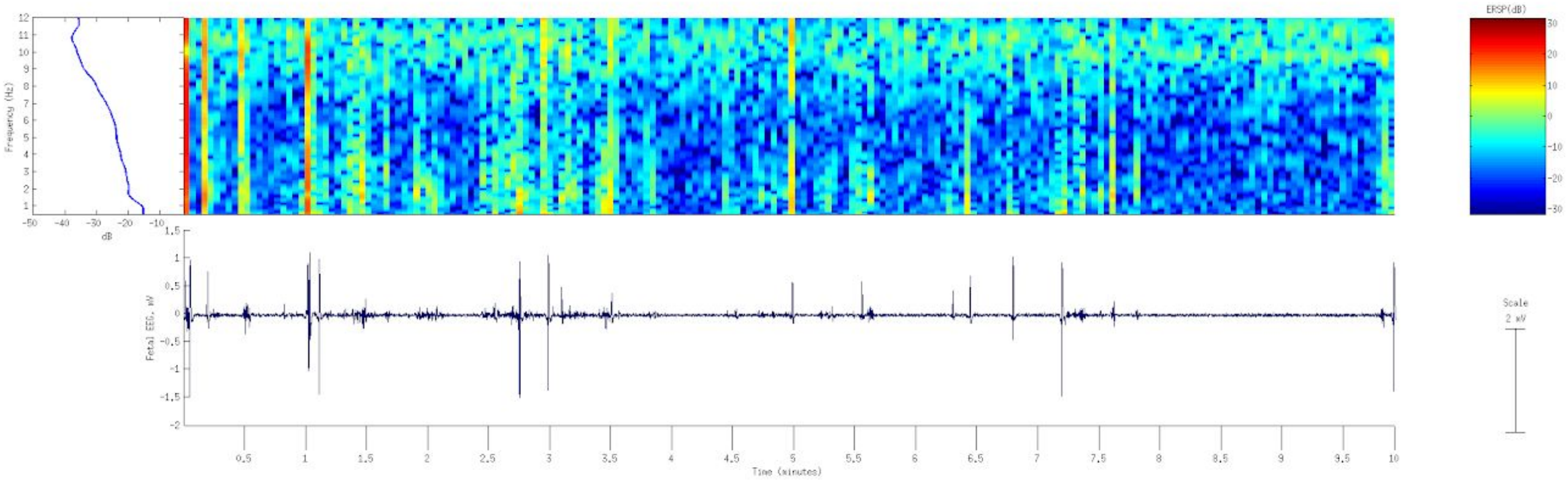

**Figure 7.** FEEG recording from the standard fetal scalp electrode during the first stage of labor. A period of ten minutes is shown with fEEG tracing (bottom) filtered 0.5-12 Hz and the corresponding power spectral analysis (top left) and wavelet transform (top left) to demonstrate the time-frequency behavior of fEEG. Note switching between delta and alpha-band activity. The X-axis shows time, with each segment corresponding to 0.5 min for a total of 10 min. Signal processing was performed in EEGLAB using Matlab 2013b, MathWorks, Mattick, MA. fEEG : fetal electroencephalogram.

The information gathered from fEEG, FHR monitoring, scalp pH measurements, Apgar score used as control measures of fetal health, and their relationships with one another were studied and the invaluable information they can provide have been demonstrated in several studies.[14,22,34,43,48,53] The 2019 Early Notification scheme progress report of the National Health System identified 70% of perinatal brain injuries as avoidable with continuous CTG monitoring.[81] Therefore, both fEEG and FHR monitoring should ideally be part of the standard of care for intrapartum surveillance allowing earlier detection of fetal distress and identification of infants at risk of abnormal neurological long term outcomes to allow timely course corrections before the irreversible injury occurs. We emphasize the intentional choice of the term "ideally" in this recommendation, as the present reality in many L&D units is that continuous EFM is not utilized and scalp electrode placement is reserved to higher-risk deliveries. We suggest that this practice is to an extent the result of disagreement about the benefits of continuous EFM or scalp electrode placement and demonstration of such benefits for prevention of brain injuries may shift the preferences toward a broader adoption of these technologies.

Animal studies, and more precisely the ones using sheep model and UCO mimicking condition of fetal ischemia have proven useful to yield better knowledge of fEEG and its usefulness as a monitoring tool during labor.[31,46,70,71,82] They are also useful in comparing treatment outcomes as shown by Gerrits et al. (2005).[77] However, while the benefit of translational medicine is indisputable, proper studies in human subjects and particularly prospective studies are still required to further establish the utility of fEEG monitoring intrapartum. Because this type of studies can be very challenging to conduct, in addition to the research aspects of fEEG, the research setting, and organization of the protocol are important for eventual success.

Below we summarize our experience with conducting a prospective fEEG study at an L&D unit (Figure 8). The study recruitment process begins with two forms of passive engagement. A potential participant's first exposure to the study is an informational flyer near the L&D reception desk. As the potential participant moves through the L&D ward, they will encounter bright purple door flyers denoting another occupant's participation in the study. Both of these engagements are low to medium impact and do not require interaction with study personnel. However, the name recognition and potential assurance of other families participating in the study lay the foundation for later direct interaction with study personnel. The next step in the recruitment process is this direct interaction. L&D staff identify potential study families and communicate the room numbers to recruiting personnel. This personnel approaches the family with an informed consent form and a summary sheet that further simplifies the objectives of the study. This step only gains initial interest from the family and is dependent on the placement of a fetal scalp electrode (FSE). If an FSE is utilized during the delivery, L&D staff will inform the technical study personnel to confirm consent and connect the study device. The device will record the data for future analysis.

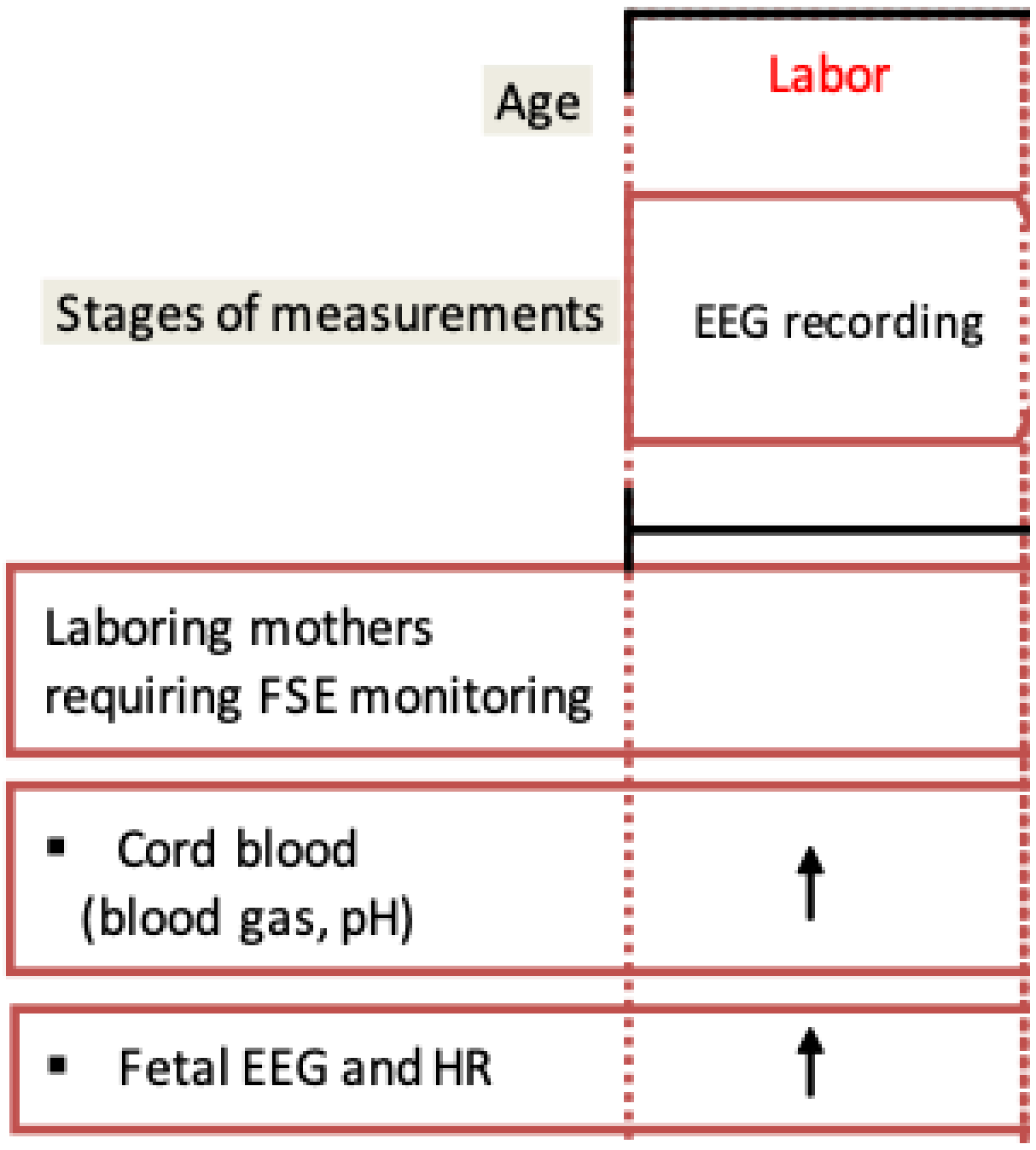

**Figure 8. Suggested study protocol.** Fetal EEG recording during labor will be followed by cord blood measurements at birth to determine the degree of acidemia and the neonatal morbidity score. FSE = fetal scalp electrode; EEG = electroencephalogram; HR = heart rate.

Obtaining clinical data for assessing the automated algorithms proved more difficult than the initial study design anticipated. Recruitment of eligible families fell well below the initial study benchmarks. We, therefore, reviewed the consent and recruitment process to better communicate the goals of collecting the necessary data. Our review determined that simplifying a study's intervention down to the required effort by families and direct impact helped cut through many potential barriers to initial participant interest. Simple solutions included accompanying a three to five bullet-point summary sheet to complement the informed consent process. When reviewing the required informed consent form, the bullet point summary helped remind a laboring mother what the study required of her family. Another way we revised our process was to be cognizant of the laboring mother's attention span and the number of hospital personnel involved with the family's clinical visit. We retrained staff to keep interactions as brief as possible. Families are inundated with rounding clinical staff across multiple shifts; a lengthy interaction with study personnel for an optional study was likely to be dismissed by a laboring mother. It is also important to avoid approaching laboring women in the second stage of labor or after a significant deceleration when they are overwhelmed.

Upon review of other studies, we discovered that this was a common mistake in subject recruitment. Often, L&D studies overlook a subject's combination of being unfamiliar with their situation, being mentally/physically overwhelmed, and being unfamiliar with the consent process. We felt that our simple mitigating processes helped increase our potential subjects' interest.

*Fetal EEG during pregnancy*

Although the focus of this review was on intrapartum fEEG studies, we cannot completely overlook the valuable information gathered from antepartum fetal magnetoencephalogram (fMEG) studies. As mentioned before, fEEG recording was first described by Lindsey (1942) in a 7-month fetus in utero and later described by Okamoto (1951) who identified EEG activity in the fetus as early as 12 weeks old.[83] In 1985, Blum et al. described a new technique, the fMEG, to record fetal brain electrical activity in utero.[84] The technique had the benefit of being non-invasive, yet allowing to obtain fMEG traces of decent quality. The technique was further perfected by Eswaran and his colleagues to minimize artifacts mainly from maternal ECG and fECG as well as from the environment.[85] The fMEG allowed to study MEG patterns associated with fetal brain maturation similar to what is seen in preterm infants.[86] The technique was also used to study behavioral states and sleep patterns associated with the gestational age allowing to gain better insight into the developing brain (Haddad 2011).[87]

As the knowledge on antepartum fMEG/fEEG expands further, it will necessarily affect the more specific intrapartum situation. Therefore, to better understand intrapartum fEEG, staying up to date on the literature related to the antepartum EEG/MEG monitoring is necessary.

## Conclusions

In this systematic review of the literature on intrapartum fEEG, we found that if a "normal" baseline EEG activity can be successfully recorded, abnormal patterns suggestive of fetal distress can also be observed. The combination of fEEG analysis with FHR monitoring as well as Apgar score can help identify patients at risk allowing early intervention. This should also help identify when the situation is not alarming, preventing unnecessary interventions such as C-section. The majority of the studies date back from the 70s with the potential that some of their data could be invalidated by newer technologies. Despite the paucity of recent studies on the subject, the over 50 years of literature on fEEG clearly demonstrates that fEEG represents a clinically tested bedside monitoring technology of fetal well-being during labor with a clear potential to detect fetal distress, complementary to FHR monitoring. FEEG intrapartum warrants prospective clinical research with modern technical capabilities of data acquisition and computerized interpretation in L&D setting.

**Conflict of Interest**

MGF holds a patent "EEG Monitor of Fetal Health" US9,215,999 and has a start-up Health Stream Analytics LLC to develop fetal EEG technologies. No other conflicts of interest are reported.

**Author Contributions**

- AC collected and screened all the studies, determined eligibility, summarized the findings from each study, and compiled them in tables. AC translated the French language manuscripts, wrote, revised, and approved the final version of the manuscript.
- YF conducted the preliminary literature review by establishing the search method and keywords and started screening some of the studies. YF contributed to the first draft and approved the final version of the manuscript.
- JF contributed to, revised, and approved the version of the manuscript.
- FK contributed to, reviewed, and approved the final version of the manuscript.
- CA contributed to the manuscript, revised the manuscript, and approved its final version.
- MGF conceived the manuscript, reviewed all studies screened, and confirmed if they were eligible or not. MGF translated the German and Russian language manuscripts, drafted, revised, and approved the final version of the manuscript.

**Funding**

MGF's work on this subject has been funded, in part, by the Western Innovation Fund (WIF), Women's Development Council, London Health Sciences Centre, London, ON, Canada; Molly Towell Perinatal Research Foundation (MTPRF), Fonds de Recherche du Québec - Santé (FRQS), Canadian Institutes of Health Research (CIHR), the University of Washington's Dept. of Obstetrics and Gynecology. We gratefully acknowledge the Perinatal Research Lab of Dr. Bryan Richardson who had supported this research in the initial stages. The referenced fetal sheep UCO models and experiments came from this lab.

**Abbreviations**

CRI: continuous rate infusion; CTG: cardiotocogram; fECoG: fetal electrocorticogram; fECG: fetal electrocardiogram; fEEG: fetal electroencephalogram; fMEG: fetal magnetoencephalogram; FHR: fetal heart rate; IUGR: Intrauterine growth restriction; PROM: premature rupture of membrane; UCO: umbilical cord occlusion.

**Tables**

**Table 1:** Summary of the 40 eligible studies and their level of evidence, number of subjects included and their gestational age.

| Authors/year | Level of evidence | Type of Study | Number of subjects | Gestational age of subjects |
|---|---|---|---|---|
| **Studies in human** | | | | |
| Thaler 2000[16] | 3 | Non-consecutive cohort study | 14 | 39.9± 1.2 weeks |
| Weller 1981[25] | 4 | Case series | 20 | Term fetus |
| Kurz 1981[30] | 3 | Cohort study | 20 | Not reported |
| Wilson 1979[43] | 2 | Inception cohort study | 25 | Full term |
| Chik 1979[78] | 5 | Author's recommendations | N/A | N/A |
| Borgstedt 1978[48] | 4 | Poor quality cohort study (biased recruitment of high-risk cases) | 158 | 40.1±2.1 weeks |
| Nemeadze 1978[45] | 3 | Non-randomized controlled cohort | 105 | N/A |
| Chik 1977[53] | 3 | Retrospective cohort study | 61 | 39.4 ± 3 weeks |
| Revol 1977[34] | 4 | Case series | 140 | 125 term fetus, 6 near-term and 9 premature between 31 and 36 weeks |
| Sokol 1977[50] | 2 | Prospective cohort study with good follow up | 38 | Not reported |
| Chik 1976[51] | 3 | Retrospective cohort study | 11 | Term fetus (mean 40.5 weeks). |
| Chik 1976[52] | 3 | Retrospective cohort study | 9 | 39.1 weeks: 7 term fetus, one 37 weeks and one 34 weeks |
| Hopp 1976[42] | 4 | Retrospective cohort study with poor follow up | 85 | Not reported |
| Borgstedt 1975[22] | 2 | Prospective cohort study with good follow up | 96 | Not reported |
| Chik 1975[28] | 3 | Retrospective cohort study | N/A | Not reported |
| Challamel 1974[33] | 4 | Case series | 100 | 92 term fetuses and 8 preterm (<36 weeks). |
| Fargier 1974[38] | 4 | Prospective cohort study with poor follow up | 120 | Not reported |
| Heinrich 1974[24] | 5 | Proof of concept | 1 | Not reported |
| Beier 1973[36] | 2 | Prospective cohort study with good follow up | 34 | Not reported |
| Carretti 1973[55] | 4 | Individual case control study | 20 | 36 to 40 weeks |
| Hopp 1973[49] | 2 | Inception cohort study | 37 | Not reported |
| Peltzman 1973[27] | 5 | Case series ≤ 5 cases | 5 | Term fetus |
| Peltzman 1973[29] | 5 | Case series ≤ 5 cases | 2 | Not reported |
| Rosen 1973[88] | 4 | Poor quality cohort study | 6 | Not reported |
| Rosen 1973[32] | 4 | Case series | 300 | Not reported |
| Chachava 1972[41] | 2 | Inception cohort study | | Not reported |
| Hopp 1972[26] | 4 | Case series | 5 | Not reported |
| Mann 1972[23] | 5 | Observation/ first principles | 50 | Not reported |
| Feldman 1970[89] | 5 | First principles | N/A | N/A |
| Rosen 1970[19] | 4 | Poor quality cohort study | 125 | Not reported |
| Chachava 1969[39] | 3 | Cohort study | 30 | Not reported |
| Rosen 1969[20] | 5 | First principles (technique description) | 14 | N/A |
| Barden 1968[37] | 4 | Case series | 6 | Not reported |
| Rosen 1965[21] | 3 | Non-consecutive cohort study | 15 | Not reported |
| **Studies in animals** | | | | |
| De Haan 1997[69] | 2 | Individual RCT | 21 | Fetal lamb: 126.5 ± 2.8 day of gestation (term 147 d) |
| Thorngren-Jerneck 2001[70] | 3 | Exploratory cohort study | 16 | Near-term fetal lambs at mean (range) gestational age 136 (134–138) days |
| Kaneko 2003[71] | 3 | Exploratory cohort study | 8 | Fetal lamb: 127 - 130 days of gestation |
| Gerrits 2005[77] | 2 | Individual RCT | 22 | Fetal lamb: 117-124 days of gestation |
| Frasch 2011[46] | 2 | Individual cohort study | 10 | Fetal lamb: 125±1 days gestation |
| Wang 2014[31] | 2 | Individual cohort study | 20 | Near term fetal lamb: 123±2 days |

**Table 2:** Summary of the method used for fetal electroencephalogram (or electrocorticogram) recording and the condition of recording for the 40 eligible studies.

| **Authors/year (ref)** | **Electrode type and placement** | **Mode/Frequency** | **Condition of recording** |
|---|---|---|---|
| **Studies in human** | | | |
| Thaler 2000 [16] | Two custom-made circular scalp EEG electrodes (suction silicone rubber cups applied by continuous negative pressure) with a central metal probe applied at the occipitoparietal or parietal region (with at least 4 cm between electrodes). FHR recorded with a scalp electrode. | Signals sampled at 250 Hz, stored and displayed by a Cerebro-trac 2500 (*SRD Medical Ltd. Shorashim, Israel*) using real-time Fourier transform (FFT) algorithm to calculate the power spectrum of fEEG. Epochs length acquisition: 4s. Band-pass filter: 1.5-30Hz. Amplifier sensitivity: 200µV. | Low risk pregnancies in the active stage of labor. |
| Kurz 1981 [90] | *N/A* | FHR, uterine contractions, fetal blood analysis and fEEG recorded polygraphically. Spectral power analysis was performed in real time sequentially and plotted in 30 s intervals continually over the course of the entire observation in waterfall style. | Normal deliveries (n=20) |
| Weller 1981 [25] | Flexible electrode minimizing ECG artifact with an incorporated guard ring surrounding the recording sites and forming the indifferent and common electrodes (acting as a short circuit to the fECG). The two electrodes are 23mm apart and inserted through a 3cm dilated cervix (after membrane rupture). | Amplifier circuit: microminiature resistors and standard low noise operational amplifier (SF C 2776UC). Input resistance: 2.5 +2.5 Ohms. Gain: 32dB. Power requirement +/- 5 volts (100µA). Noise level set at <2µV peak to peak (1 to 40Hz). Infrared telemetry used to convey the fEEG to display and record equipment. | Monitoring during the second phase of uncomplicated labor in primigravid mothers with term fetus under epidural anesthetic. Recording in a standard delivery room. |
| Wilson 1979 [43] | 8 channel portable Elena Schonander Recorder. | Continuous recording. Fetal EEG analyzed in 10s epochs. | High-risk African primigravida mother. |
| Chik 1979 [78] | *N/A* | *N/A* | *N/A* |
| Borgstedt 1978 [48] | Same as Rosen 1973b[32] | Same as Rosen 1970 | Selected high risk cases (based on prenatal maternal complication or suspected intrapartum fetal distress) |
| Nemeadze 1978[45] | Simultaneous fetal EEG and ECG recording from fetal head from the moment of the first stage of labor when cervical dilation was 4-6 cm. No further details provided. | N/A | Study of the impact of premature rupture of the membrane in a group of healthy women (n=60) and a group of women with mild nephropathy (n=45). fEEG and fECG are recorded simultaneously during labor and correlated to onset of PROM and the Apgar score. |
| Chik 1977 [53] | 2 scalp cup electrodes held by applied suction on fetal head with a central silver or platinum pin avoiding penetration of the fetal skin. | Same as Rosen 1970 | File selection of high-risk infants monitored for at least 1 hour during labor. |
| Revol 1977 [34] | 2 scalp cup electrodes held by continuous suction with a central silver pin, placed on the parietal region of the skull. | Continuous recording during labor and after. EEG device: Mingograf EEG 8 Siemens (ink jet print). Filter from frequencies >30Hz. Band-pass with 0.15s time constant. Calibration 1s, 50µV. | 125 cases: normal labor (25), abnormal labor (18), fetal distress (26), under anesthesia (alfatesine n=21) or ketamine (n=26). |
| Sokol 1977[50] | Same as Rosen 1973b PMID: 4681833 | Same as Rosen 1970 | EEG recorded in suspected increased risk deliveries. EEG findings assessed in relation to follow up at one year. |
| Chik 1976a [51] | Same as Rosen 1973b | Same as Rosen 1970 | Same as Rosen 1973b |
| Chik 1976b [52] | Same as Rosen 1973b | Same as Rosen 1970 | Same as Rosen 1973b |
| Hopp 1976[42] | Same as Hopp 1972[26] | Same as Hopp 1972 (25) | Simultaneous recording of fEEG and CTG in 220 fetuses, (85 cases ultimately included). with some under conditions of intermittent hypoxia due to uterine contractions and after maternal administration of drugs. |
| Borgstedt 1975 [22] | 2 scalp cup electrodes held by applied suction on the fetal head with a central silver or platinum pin avoiding penetration of the fetal skin. Electrodes implanted once cervix dilation reached 2 cm. | Same as Rosen 1970 | Patients selected because of increased prenatal risk or suspected fetal distress during labor. |
| Chik 1975[28] | Same as Rosen 1973b | Same as Rosen 1970 | Retrospective evaluation of fEEG recording. |
| Challamel 1974 [33] | 3 Modified Dassault electrode (cup with central pin) placed on the scalp with one placed in parietal position and a minimal distance between electrodes of 4 cm. | EEG device: Mingograf EEG 8 Siemens (ink jet print). Filter from frequencies >30Hz. Band-pass with 0.15s time constant. Calibration 1s, 50 µV. Continuous recording (30 min to 5h intrapartum). Best derivation exploited. | Monitoring during different conditions of labor (including fetal distress). |
| Fargier 1974[38] | Same as Challamel 1974 | Same as Challamel 1974. No filter <70Hz. | Monitoring during different conditions of labor and different drug administrations. |
| Heinrich 1974[24] | Same as Rosen 1973[32] | Intrapartum multimodal fetal monitoring device: RFT Fetal Monitor BMT-504. | One example of EEG recorded in a neonate to show how the new monitor can be used. |
| Beier 1973 [36] | 1 cup electrode consisting of a central needle pin surrounded by a 4 cm disc and two suction grooves connected with a suction device to ensure firm electrode placement over the fetal skull. | 1-channel EEG; simultaneous recording of fetal ECG, FHR and intraamniotic pressure channels. Calibration 1s, 50μV. | About 30 min duration intrapartum monitoring in 34 fetuses from healthy pregnancies, labor and postnatal outcomes. In 15 fetuses/neonates, the corresponding postnatal recordings were also made. |
| Carretti 1973[55] | Plexiglas suction cup with 6 electrodes around its periphery placed on the fetus occiput following >4cm cervix dilation. | Galileo apparatus allowing multipolar EEG recording. | Comparison of fEEG in healthy mother before and after oxygen administration. |
| Peltzman 1973 [27] | 2 flexible stainless-steel screw electrodes (impedance less than 400 Ohms in all cases) | FEEG is continuously monitored on a polygraph (Grass instrument Co., Quincy, Massachusetts). Band-pass filter: 1.0 to 35.0Hz. A PDP-7 computer allows real time analysis and storage of the fEEG before the information is transmitted to a 64-channel analogue/digital converter. The fEEG for each 5s epochs is set to zero mean. Mean fEEG and time integrated fEEG amplitude are also computed. The program then computes a zero line-cross count on the zero-mean fEEG by checking along the wave within each 5 sec epoch and recording the line cross each time the polarity changes. Data based upon 12 to 19 continuous 5 s epochs are presented as graphs with plotted points from the beginning of the recording and represent artifact-free analogue fEEG. | Uneventful labor and delivery (3 out of 5 were induced labor). Effect of analgesia with meperidine (n=3) and diazepam (n=2) or paracervical block (n=2) evaluated. |
| Peltzman 1973 [29] | 2 flexible stainless-steel bipolar screw electrodes placed 3-4cm apart of the parietal area of the fetal scalp. | Similar to Peltzman 1973 | Uneventful labor with paracervical block (1% mepivacaine) administered |
| Rosen 1973a [14] | Same as electrodes as Rosen 1970Electrodes placed along the sagittal suture and between the fontanelles to avoid the forceps blade in forceps birth (over the parietal region for spontaneous birth) | Similar to Rosen 1970 | FEEG during expulsion efforts and during low forceps deliveries are compared to EEG during spontaneous deliveries. Pre-forceps EEG also recorded for comparison. |
| Rosen 1973b [32] | 2 electrodes applied over the parietal and consisting of a silver pin in the center of a lucite disc maintained in place by continuous suction after application with an interelectrode distance of at least 4 cm. | Similar to Rosen 1970 | EEG recording during different labor situations (eventful vs complicated labor in neurologically abnormal infants). |
| Hopp 1973[49] | Same as Hopp 1972[26] | Same as Hopp 1972[26] | Simultaneous recording of fetal EEG, fECG and CTG in 37 fetuses during labor in the late first stage and in the stage of active pushing (second stage). |
| Chachava 1972[41] | Two fEEG and fECG electrodes placed onto the fetal head, as remotely from each other as possible, fixated using vacuum suction. | N/A | fEEG, fECG and maternal EEG recorded following cervical dilation and rupture of membranes during the first stage of labor. |
| Hopp 1972[26] | 3 cup electrodes made of a 40mm silver disc with a 5mm center pin are used: two biparietal and one on midline, placed and held by suction. | Fetal ECG, CTG and intra-amniotic pressure acquired with an 8-channels EEG device. Artifact free fEEG was achieved when bimodal rejection mode was chosen (i.e., biparietal EEG electrode against a ground). | Simultaneous recording of fEEG and CTG during different labor conditions including normal labor and cardiac decelerations. |
| Mann 1972 [23] | Two silver disc electrodes consisting of a vacuum contact cup, tube and wire connection and a vacuum electric plug to the vacuum module are placed with membranes ruptured after cervix dilation reaches 3-4cm. The electrodes are placed up against the vertex about 1 to 2 cm apart. | Serial bipolar EEG are obtained by direct write out on the dual-channel San'Ei electroencephalograph (San'Ei Instrument Co., Div. of Medical System Corp, Great Neck, New York) | Description of the electrode used to record fetal EEG during labor (50 recordings). Specific conditions of recording not documented. |
| Rosen 1970 [19] | Two electrodes consisting of a platinum needle embedded in the center of a lucite disc with the firm margin of the disc preventing deep penetration of the needle. Continuous suction is applied within the disc to draw the scalp up to the needle recording point (with possible skin penetration of 1-2mm). Circular grooves into the lucite disc prevent the skin from occluding the suction. Electrodes implanted as soon as cervical dilation reached 3cm over the parietal areas. | Bipolar EEG recording using an 8-channel Dynograph Recorder run at 30mm/s. Time constants are arranged to allow recording wave frequencies between 0.5 and 32 Hz/min. Recording amplitude of 20µV/cm and 50µV/cm are used. | 125 fetuses recorded during different conditions (normal labor, forceps assisted delivery and following drug administration). |
| Chachava 1969[39] | Two fEEG scalp electrodes were held with vacuum suction and placed 2-3cm apart on the fetal head (after 3-4 cm cervical dilation and ruptured membranes). | Bipolar EEG recorded with either a 4-channel locally made device or an 8-channel EEG device by Orion. | fEEG recorded intrapartum: 20 with normal labor and 10 with complications. |
| Rosen 1969 [20] | Two electrodes made of an outer shell in silicone rubber with its periphery circumscribed by a silicon rubber guard ring impregnated with powdered silver (the outer ring is a patient ground) and a platinum needle sheathed in a Teflon tube soldered to a silver plate wire with its distal 2mm left bare, are used. The electrodes are introduced after membranes ruptured once cervix dilation reached 3 cm. Suction is turned on after the electrode is applied on the fetal head. | Recording in standard EEG fashion with filters admitting wave frequencies between 0.5 and 32 Hz. Paper speed is 30 mm/s recording amplitude at 20 to 50µV/cm. To document the EEG activity as brain waves, the technique of evoked response to a 34dB, 35 ms sound (repeated every 4s) is used. | Fetal EEG during labor (no more detail provided). |
| Barden 1968 [37] | Skin-clip electrode placed on the presenting vertex. | A summing computer (Mnemotron Corporation CAT-Model 400B. is used to accentuate fEEG response time locked to an acoustic signal and to cancel non time-locked fECG and random electrical noise signals. FEEG responses were sequentially averaged. | Elective induced labor. FEEG recorded before, during and after the onset of a 1000 Hz, pure tone of 450s duration (88 to 105dB). |
| Rosen 1965 [21] | Metal skin clips soldered to shielded cable, coated with non-conductive plastic glue and filed at their tip to prevent deep scalp penetration and attached to the vertex with a modified uterine packing forceps. Mother grounded to the machine by a strap around the thigh. | Grass Model III portable EEG (Grass Instrument Co., Quincy, Mass). | Normal labor condition studied. |
| **Studies in animals** | | | |
| De Haan 1997 [69] | Two pairs of EEG electrodes (AS633-5SSF, Cooner Wire Co., Chatsworth, CA) placed on the parasagittal fetal dura through burr holes (skull coordinates relative to bregma: anterior 5 mm and 15 mm, lateral 10 mm). | The total fEEG intensity is median filtered to remove short-term (<20 min) fluctuations, log transformed to get a better approximation of the normal distribution and normalized with respect to the 12-h baseline. Total EEG intensity, EEG spectral edge (upper 90% of frequency), and cortical impedance are measured in 15-min periods during UCO and in 5-h intervals after the last occlusion. Epileptiform activity and spike detection software (Monitor, Stellate Systems, Quebec, Canada) is used to scan the raw EEG file. | FEEG recorded after UCO with fetuses randomized to one of three groups: group I, repeated total UCO for 1 min every 2.5 min; group II, repeated total UCO for 2 min every 5 min; and group III, no occlusions (sham controls). UCO is repeated until fetal arterial blood pressure had fallen below 2.7 kPa (20 mm Hg) during two successive occlusions, or until fetal blood pressure failed to recover to baseline levels when the next occlusion is due. |
| Thorngren-Jerneck 2001 [70] | Two EEG electrodes (shielded stainless steel) placed bilaterally over the parietal cortex (10 mm anterior of bregma and 15 mm lateral of midline), inserted through drilled holes in the parietal bone. A subcutaneous reference electrode is placed posteriorly in the midline of the skull. | No detail provided. | 16 near-term fetal lambs: 8 lamb fetuses exteriorized and subjected to total UCO in a water bath, four lamb fetuses exteriorized and serving as sham controls and four lamb fetuses immediately delivered after minimal preparation and serving as healthy controls |
| Kaneko 2003 [71] | Electrodes of Teflon-coated stainless-steel wire (Cooner Wire, Chatsworth, Calif) implanted biparietally on the dura for recording of electrocortical activity. | No details provided. | ECoG recording following repeated UCO for 4 minutes, every 90 minutes, and over 6 hours (total 4 UCO). |
| Gerrits 2005 [77] | Two pairs of EEG electrodes (AS633-5SSF; Cooner Wire Co., Chatsworth, CA) placed on the dura over the parasagittal parietal cortex (5 and 15 mm anterior and 10 mm lateral to the bregma), with a reference electrode sewn over the occiput. | Fetal parietal EEG and impedance recorded continuously. Signals are averaged at 1-min intervals and stored to disk by custom software (Labview for Windows; National Instruments Ltd, Austin, TX), running on an IBM compatible computer. The EEG signal is low pass filtered at 30 Hz, and the intensity spectrum and impedance signal are extracted. The raw EEG signal is recorded for off-line detection of seizure events. | Fetal lamb subjected to selective cooling of the head following cerebral ischemia (with one control group) |
| Frasch 2011[46] | Stainless steel ECoG electrodes are implanted biparietally on the dura through small burr holes in the skull bone  placed ~1–1.5 cm lateral to the junction of the sagittal and lambdoid sutures. The bared portion of the wire to each electrode is rolled into a small ball and inserted into each burr hole to rest on the dura with a small plastic disk covering each burr hole held with tissue adhesive against the skull bone. A reference electrode is placed in the loose connective tissue in the midline overlying the occipital bone at the back of the skull. | ECG and ECoG are recorded and digitized at 1000 Hz. For ECG, a 60 Hz notch filter is applied.  For ECoG, a band pass 0.3–30 Hz filter is used. The ECOG signal is sampled down to 100 Hz prior to analysis. Voltage amplitude and 95% spectral edge frequency (SEF), are calculated over 4 s intervals (Spektralparameter, GJB Datentechnik GmbH, Langewiesen, Germany). | Fetal lamb studied after series of mild, moderate and severe UCO until fetal arterial pH fell below 7.00 |
| Wang 2014 [31] | Fetal instrumentation after exteriorization: a modified FHR electrode with a double spiral placed on the fetal head is used. | A PowerLab system is used for data acquisition and analysis (Chart 5 For Windows, AD Instruments Pty Ltd, Castle Hill, Australia). For fEEG recording, a band pass 0.3–30 Hz filter were used. Prior to analysis, fECoG and fEEG were sampled down to 100 Hz. | FEEG (and fECoG) recorded in near term fetal lamb during repeated UCO. |

fECG: fetal electrocardiogram; fECoG: fetal electrocorticogramm, fEEG: fetal electroencephalogram; UCO: umbilical cord occlusion

**Table 3.** Summary of the possible findings in fetal EEG tracings, as reported in the different studies, under different labor conditions.

| **Fetal EEG conditions** | **Observations** |
|---|---|
| Baseline fEEG during normal labor | **Rosen 1965:** Low voltage baseline pattern with a low voltage (20 µV), faster frequency (8 Hz) after umbilical cord clamping. On most tracings, the electrical activity before and after the first breath and before and after umbilical cord clamping did not appear to change abruptly. FEEG activity recorded early in labor has a baseline pattern similar to that of the alert neonate.[21]<br><br>**Chachava 1969:** Low amplitude waves (10-30 µV) of 0.04-2 s in duration with observation of alpha, beta, theta and delta waves.[39]<br><br>**Rosen 1969:** The fEEG voltages vary from 5 to 50µV per cm with waves frequencies between 1 and 25 Hz.<br><br>**Rosen 1970:** The wave frequencies vary between 0.5 and 25 Hz with the predominant frequencies in the 2.5-5 Hz. Patterns similar to those present in neonates of the same birth weight.[35]<br><br>**Hopp 1972:** Amplitude of fEEG ranges between 10 and 70 µV. Frequency varies considerably and ranges between 2 - 20 Hz.[26]<br><br>**Mann 1972:** Rhythm consisting of 1 to 3 Hz waves with an amplitude of about 40 to 75 µV and superimposed faster frequencies of 4 to 8 Hz and 10 to 30 µV.[23]<br><br>**Fargier 1974:** Similar to that of neonates of same gestational age with presumptive alternance of awake/sleep states (both deep and active sleep).[91]<br><br>**Borgstedt 1975:** Wave frequencies of 0.5 to 25 Hz with an amplitude generally between 50 and 100 µV/cm similar to neonatal EEG.[22]<br><br>**Thaler 2000:** Two fundamental EEG patterns are identified: high voltage slow activity (HVSA)(quiet behavioral state) and low voltage fast activity (LVSA) (active behavioral state). On average, LVSA was present 60.1% of the time and HVSA was present 39.9% of the time.[16] |
| Contractions | **Mann 1972:** No fEEG changes (even with very intense oxytocin-induced contractions).[23]<br><br>**Rosen 1973:** No fEEG changes even with stronger contractions during the second stage of labor.[3214]<br><br>**Beier 1973:** No fEEG patterns associated with contractions in most cases, but in some fetuses, a 20 - 40 s delayed increase in amplitude which normalized 5-10 s after contractions ended is observed.[36]<br><br>**Hopp 1976:** fEEG shows reduction of frequency and increase in wave amplitude during contractions.[42,80]<br><br>**Chachava 1972, Fargier 1974, Revol 1977, Weller 1981:** No fEEG changes with normal contractions.[25,34,38,41] |
| Spontaneous birth | **Rosen 1973b:** Low voltage irregular activity. Artifactual distortion of the fEEG baseline characterized by large rolling waves of almost 2s in duration due to electrodes movements when the vertex moves rapidly and the fEEG is recorded in the microvolt range.[32] |
| *Abnormal situations* | |
| Fetal heart rate (FHR) decelerations | **Rosen 1970:** Previously recorded higher voltage fEEG pattern abruptly changed in association with depression of the FHR to an almost flat or baseline pattern. The tracing returned to pre-existing patterns after FHR returned to normal. This change was found to be most commonly associated with delayed FHR decelerations.[35]<br><br>**Hopp 1972**: Fetal bradycardia, especially during contraction-associated late decelerations, was accompanied by reduction in fEEG waves (lower frequency) and occurrence of fEEG spike potentials.[26]<br><br>**Rosen 1973b:** No change with early decelerations. With variable and late deceleration, the fEEG appeared to lose the faster rhythms, followed by a more apparent slowing. Then, isoelectric to almost flat periods with rare bursts of fEEG are seen and finally a totally isoelectric interval might be observed sometimes for longer than 10s (rarely more than 30s). As the FHR returns to its baseline rate, the reverse of this progression takes place. The entire sequence from onset to return may last from 30s to longer than one minute.[32]<br><br>**Fargier 1974:** No changes with early deceleration.[38]<br><br>**Revol 1977:** Early deceleration was only associated with fEEG changes only with FHR below 90 bpm.[34]<br><br>**Hopp 1976:** During severe variable decelerations, fEEG showed waves of low amplitude and near isoelectricity and intermittent spike potentials between contractions.[42,80]<br><br>**Wilson 1979:** A significant relationship was noted between the increasing percentage of electrocerebral silence and the development of FHR deceleration patterns during labor. Early FHR deceleration was also associated with prolonged silence in the fEEG.[43] |
| Tachycardia | **Rosen 1970:** fEEG changes consistent with voltage suppression, i.e., generalized decrease in the wave amplitude of a constant nature often associated with increasing intervals of flattening without EEG activity.[35]<br><br>**Thaler 2000:** FHR accelerations typically associated with periods of low voltage slow activity.[16] |
| Fetal distress | **Chachava 1969:** fEEG of a baby born asphyxiated and demised within 15 min postpartum showed fast activity around 6 Hz that may represent brain hypoxia. High amplitude low-frequency waves were suspected to be signs of brain injury during labor.[39]<br><br>**Hopp 1973:** Patterns pathognomonic for abnormal fEEG and suspected fetal brain injury: 1) extremely high voltage activity (> 80μV), 2) extremely low voltage activity(< 10μV), 3) spike potentials as a sign of epileptiform activity, 4) bihemispheric differences, 5) Reduction of fEEG frequency during pathologically silent FHR pattern.[49]<br><br>**Rosen 1973b:** Non-transient fEEG changes such as sharp waves were defined as repetitive waves, always of the same polarity, generally higher in amplitude than the surrounding fEEG and generally less than 50 ms in duration. When observed, they were usually present at the onset of recording and continued throughout labor. These sharp waves seemed to be more frequent in the developmentally abnormal child at one year of life. The combination of sharp waves and low voltage did not occur in the normal population. Therefore, this type of activity may suggest fetal distress.[32]<br><br>**Borgstedt 1975:** Isolated sharp waves were more frequent in newborn with abnormal neurologic findings than in those neurologically normal.[22]<br><br>**Revol 1977:** Cases with the combination of abnormal fEEG, abnormal cerebral blood flow, low pH (<7.25) and abnormal Apgar score had the lowest Apgar score. FEEG did not always normalize after in utero resuscitation despite the correction of FHR and pH.[34]<br><br>**Borgstedt 1978:** Prolonged voltage suppression (<20μV) (during the entire recording or for at least several seconds, alternating with normal trace and continuing through birth) is correlated with lower 1min- and 5min- Apgar score and the need for post-partum resuscitation.[48]<br><br>**Wilson 1979:** Significant correlation between the development of electrocerebral silence in the fEEG and the development of fetal acidosis. The rapid deterioration in fEEG occurred as the pH fell and even at preacidotic levels (pH of 7.2 to 7.25) marked changes were present with the cessation of electrical activity in the fetal brain.[43]<br><br>**Nemeadze 1978**: No effect of PROM on fEEG, but pregnancy history compounds fEEG response to PROM as follows. The brief fEEG period in response to a PROM event including the immediate 5-min post-PROM represents an adaptive functional response of the fetal brain indicative of fetal reserve. Under normal delivery, this response subsides within 1-3 min post-PROM and is associated with a healthy birth (in the absence of any other delivery complications). In contrast, prolongation of this |
| | recovery period post-PROM to 4-5 min indicates a reduction in fetal adaptation capability and was associated with brain injury at birth.[45] |
| Head compression from cephalopelvic disproportion. | **Wilson 1979:** Head compression did not appear to influence fetal brain activity.[43] |
| Uterine hypertonia or hyperkinesis | **Challamel 1974, Fargier 1974, Revol 1977:** Decreased activity and flattening of the fEEG signal.[33,34,38] |
| Forceps | **Rosen 1973a:** Forceps application was not associated with any change, but traction was, with an almost flat fEEG tracing observed.[8814]<br><br>**Challamel 1974, Fargier 1974:** High forceps extraction compared to low forceps extraction was always associated with fEEG changes during the traction phase and characterized by flattening of the trace returning to normal after a few seconds if the extraction was short and not too intense. Repeated and prolonged tractions were associated with persistent isoelectric trace up to the birth of the child with changes persisting for 20 min after.[33]<br><br>**Revol 1977:** During forceps or vacuum extraction, fEEG changes that disappear just before the expulsion effort or persistent fEEG changes sometimes to the point of isoelectric trace were noted in all but one case.[34] |
| Oxygen administration to the mother | **Carretti 1973:** FEEG changes within 1.5 - 2 minutes after initiation of $O_2$ are characterized by a progressive increase in amplitude and frequency of the waves (from 1-5 Hz to 8-12 Hz) reaching a maximum at 7-8 min followed by a decrease in the activity of the trace to return to baseline activity after 12-15 minutes in half of the cases.[55] |
| *Drugs* | |
| Meperidine | **Rosen 1970:** Early responses: a transient increase in delta wave frequencies (2.5-5 Hz), about 50 μV in amplitude first seen between 1 and 2 minutes after IV injection of the drug followed by trace alternance-like pattern of bursty activity with 5 min after the mother was given the medication. This pattern could last as long as 2 h after the injection. As the time interval after injection increased, the presence of faster, lower voltage forms (5-10 μV) and (15-25 Hz) in the beta range became more obvious.[19,35]<br><br>**Peltzman 1973a:** No identified fEEG changes.[27,29] |
| Ketamine | **Fargier 1974:** Development of sharp theta activity on an initially normal baseline, then progression to fewer waves and flattening of the trace to the point of isoelectricity with occasional bursts of theta activity.[38] |
| Pethidine | **Hopp 1976:** One min post injection of a 50 mg dose, there is a reduction of amplitude and frequency of fEEG activity. These changes are more pronounced 4 min post injection. At 6 min post injection, resynchronization is observed. These effects persisted for 25 min post injection and fEEG normalized more or less within 105 min post injection.[42,80] |
| Barbiturate | **Fargier 1974:** Sodium thiopental: same as ketamine as well as small high-frequency low voltage waves on a normal baseline progressing to decreased activity and flattening of the trace.[38]<br><br>**Revol 1977 :** Significant changes with long periods of isoelectric traces.[34] |
| Local anesthesia | **Rosen 1970:** With local carbocaine, transient increase in higher voltage (50µV/cm) bursty waves (15-25 Hz) was noted. These changes appeared to be transient.[19,35]<br><br>**Peltzman 1973a and 1973b:** Decrease in fEEG amplitude.[27,29]<br><br>**Challamel 1974:** Local epidural with marcaine or bupivacaine associated with a high-frequency rhythm (when using a filter frequency >30Hz) with clusters of rhythmic theta waves.[33] |
| Penthrane (anesthetic gas) | **Rosen 1970:** Trace alternant picture persists while the gas is being administered during the terminal stages of labor.[19,35] |
| Diazepam | **Peltzman 1973a:** No identified fEEG changes.[27,29]<br><br>**Hopp 1976:** The fEEG frequency decreased within 30 min post injection of 10mg of diazepam and the amplitude increased to 80 µV. The EEG was normal when recorded in the neonate 40 min after the injection.[42,80] |
| *Less mature infant with analgesic medications* | **Rosen 1973b:** Persistence of all voltages below 20 µV with prolonged intervals of isoelectricity (low voltage tracing) associated with an initially normal amplitude and pattern of recording that then changes to persistent low voltage with prolonged periods of isoelectricity.[32] |
| *Long-term outcome* | |
| Infants with normal long-term outcome | **Chik 1976a:** Study of fEEG of children neurologically normal at 1 year of age: the mixed pattern was predominant accounting for 41.2% of the epochs. Trace alternants accounted for 32.2%, high voltage slow for 21.5 % and low voltage irregular for 4.4% of the patterns. Less than 0.2% showed depression or isoelectricity. In the neonatal EEG studied, there was a decrease in the relative frequency of mixed and an increase in high voltage slow patterns.[51,52] |
| Infants with abnormal long-term outcome | **Borgstedt 1975:** Sharp waves that appear in isolation and not as part of burst activity were identified as abnormal.[22]<br><br>**Chik 1976b:** Study of fEEG of children neurologically abnormal at 1 year of age: the relative frequency of the low voltage irregular pattern is increased with a consequently decrease in mixed and high voltage slow patterns. The mean relative frequency of the low voltage irregular pattern was significantly greater with the lower Apgar scores (less than 9). Therefore, low voltage irregular activity occurred more frequently in the neurologically abnormal group (compared to the neurologically normal group).[51,52]<br><br>**Sokol 1977:** The detection of the presence of sharp waves alone on fEEG allowed to correctly classify 76% of patients in terms of the neurological outcome at 1 year (normal vs abnormal). Detection of voltage depression alone appropriately identified 68% patients. The observation of a combination of sharp waves and/or prolonged voltage depression improved the identification of abnormal infants but misclassified half of the normal infants as abnormal (66% correct). These results confirm the relationship between sharp waves and voltage depression in the fEEG and abnormal infant outcome.[50] |

bpm: beat per minute, EEG: electroencephalogram, fEEG: fetal electroencephalogram, FHR: Fetal heart rate; Hz: Hertz, s: seconds.